\documentclass[a4paper]{article}
\usepackage{amssymb}
\usepackage{amsmath}
\usepackage[a4paper]{geometry}
\usepackage[onehalfspacing]{setspace}
\usepackage{numinsec}
\usepackage{amsfonts}
\usepackage{amstext}

\newtheorem{theorem}{Theorem}
\newtheorem{lemma}{Lemma}
\newenvironment{proof}[1][Proof]{\noindent\textbf{#1.} }{\ \rule{0.5em}{0.5em}}
\input{tcilatex}
\begin{document}

\title{{\LARGE Testing Multiple Inequality Hypotheses :} \\
{\LARGE A Smoothed Indicator Approach}\thanks{%
This paper is a substantial revision of Chapter 3 of the first author's
doctoral dissertation (Chen 2009) and the subsequent Cemmap working paper
(Chen and Szroeter 2009). We thank Oliver Linton, Sokbae Lee, Yoon-Jae
Whang, Chungmin Kuan, Hidehiko Ichimura and Joon Park for helpful comments.
We are also grateful to seminar participants for various insightful
discussions of this work presented in 2009 International Symposium on
Econometric Theory and Applications and 2010 Royal Economic Society Annual
Conference.}}
\author{Le-Yu Chen\thanks{%
Corresponding author. E-mail : lychen@econ.sinica.edu.tw} \\
Institute of Economics, Academia Sinica \and Jerzy Szroeter\thanks{%
E-mail : j.szroeter@ucl.ac.uk} \\
Department of Economics, University College London}
\date{Revision : June 2012}
\maketitle

\begin{abstract}
{\normalsize This paper proposes a class of origin-smooth approximators of
indicators underlying the sum-of-negative-part statistic for testing
multiple inequalities. The need for simulation or bootstrap to obtain test
critical values is thereby obviated. A simple procedure is enabled using
fixed critical values. The test is shown to have correct asymptotic size in
the uniform sense that supremum finite-sample rejection probability over
null-restricted data distributions tends asymptotically to nominal
significance level. This applies under weak assumptions allowing for
estimator covariance singularity. The test is unbiased for a wide class of
local alternatives. A new theorem establishes directions in which the test
is locally most powerful. The proposed procedure is compared with
predominant existing tests in structure, theory and simulation.\bigskip }

\textbf{KEYWORDS} : {\normalsize Test, Multiple inequalities, One-sided
hypothesis, Composite null, Binding constraints, Asymptotic exactness,
Covariance singularity, Indicator smoothing\bigskip }

\textbf{JEL SUBJECT AREA} : {\normalsize C1, C4}

\newpage
\end{abstract}

\section{Introduction\label{nor_s1}}

This paper is concerned with the problem of testing the null hypothesis $%
H_{0}$ that the true value of a finite $p$-dimensional parameter vector $\mu 
$ is non-negative versus the alternative that at least one element of $\mu $
is strictly negative. A major problem for testing such hypotheses has been
dependence of null rejection probability on the unknown subset of binding
inequalities (zero-valued $\mu _{j}$). Under $H_{0}$, the asymptotic
distribution of a nontrivial test statistic is typically degenerate at
interior points (all elements of $\mu $ strictly positive) of parameter
space. But at boundary points (one or more elements zero), that distribution
is non-degenerate and may depend on the number and position of the zero
elements but not on strict positives. In consequence, determining the
critical value to be used for the test at some nominal significance level $%
\alpha $ is a nontrivial issue. The classic least favorable configuration
(LFC) approach seeks the parameter point in the null that maximizes the
rejection probability (e.g., see Perlman (1969) and Robertson, Wright and
Dykstra (1988)). This principle risks yielding tests which have
comparatively low power against sequences of alternatives converging to
boundary points which are not LFC. To improve test power, recent literature
has proposed using data-driven selection of the true binding inequalities in
place of the LFC point to compute test critical values. Whatever the
critical value, it is important to demonstrate that null rejection
probability does not exceed $\alpha $ uniformly over all $H_{0}$-compliant
data generating processes for sample size large enough. Such uniformity has
been emphasized in recent literature (e.g., see Mikusheva (2007), Romano and
Shaikh (2008), Andrews and Guggenberger (2009), Andrews and Soares (2010)
and Linton et al.\thinspace (2010)) to ensure validity of asymptotic
approximation to actual finite sample test size especially when the test
statistic has a limiting distribution which is discontinuous on parameter
space. Regardless of whether the binding inequalities are fixed according to
the LFC or determined via a stochastic selection mechanism, the functional
forms of test statistics proposed in this literature are generally
non-smooth and hence computation of test critical values requires simulation
or bootstrap.\medskip

The contributions of the present paper are as follows. We develop a multiple
inequality test whose implementation does not require computer intensive
methods. The central idea is to construct a sequence of origin-smooth
approximators of indicators underlying the sum-of-negative-part statistic
for testing multiple inequalities. The approximation is a form of indicator
smoothing in the spirit of Horowitz (1992), enabling standard asymptotic
distribution results and obviating simulation and bootstrap computation of
test critical values. Moreover, the test allows for estimator covariance
singularity.\medskip

The test statistic of this paper has a non-degenerate asymptotic
distribution of simple analytic form at boundary points of the null
hypothesis but becomes degenerate at interior points. Despite this type of
discontinuity, the test critical value can be fixed ex ante without
compromising asymptotic validity in the uniform sense that the limit of
finite sample test size (defined as supremal rejection probability over all $%
H_{0}$-compatible data generating processes) is equal to the nominal size.
We prove that this uniformity property holds for every approximator in a
wide class allowed by the paper.\medskip

The smoothing design of this paper embodies a data driven weighting scheme
which automatically concentrates the test statistic onto those parameter
estimates signaling binding inequalities. This feature is connected to
methods of binding inequality selection used in Hansen (2005), Chernozhukov
et al.\thinspace (2007), Andrews and Soares (2010) and Linton et
al.\thinspace (2010). Indeed, the smoother can also be interpreted as an
asymptotic selector and the key component of our test statistic coincides
with the sum of elements of the difference between the estimated and
recentered null-compatible mean used to obtained the simulated test critical
values for Andrews and Soares (2010)'s generalized moment selection (GMS)
based tests. The difference itself, however, is not within the class of test
statistics covered by the theory of these authors but its properties emerge
from the theory developed in the present paper.\medskip

The relative computational ease of the test of this paper might be expected
to carry a cost in terms of power. However, as we show, the test is
consistent against all fixed alternatives and is unbiased for a wide class
of local alternatives. In comparison with existing tests, its relative
strength varies with the particular direction of local alternative. We
provide a new theorem establishing directions in which the test is locally
most powerful. Monte Carlo results support the theory and reveal that finite
sample performance of the present test is not dominated by the GMS based
tests. \medskip

We now review relevant test methods in addition to the works cited above.
The QLR test has been well developed in the inequality test literature. See,
e.g. Perlman (1969), Kodde and Palm (1986), Wolak (1987, 1988, 1989, 1991),
Gourieroux and Monfort (1995, chapter 27) and Silvapulle and Sen (2005,
chapters 3-4). This test is also applied in the moment inequality literature
(see Rosen (2008), Andrews and Guggenberger (2009) and Andrews and Soares
(2010)). The asymptotic null distribution of the QLR test statistic
generally has no analytical form. Since computing this test statistic
requires solution of a quadratic optimization program subject to
non-negativity constraints, simulation and bootstrapping for the test
critical value is particularly heavy. \medskip

An extreme value (EV) form of test statistic was developed by White (2000)
in the context of comparing predictive abilities among forecasting models.
Such a statistic is lighter on computation but its asymptotic null
distribution remains non-standard. Hansen (2005) incorporates estimation of
actual binding inequalities to bootstrap null distribution of the extreme
value statistic. Hansen's refinement is a special case of the GMS based
critical value estimation proposed by Andrews and Soares (2010) who also
consider a broad class of test functions including both the QLR and other
simpler forms using negative-part functions. \medskip

The rest of the paper is organized as follows. Section 2 summarizes the
method of Andrews and Soares (2010) for testing with estimated critical
values which embody the GMS procedure for estimation of binding
inequalities. We contrast that with the smoothing approach of this paper and
highlight connecting features. Section 3 sets out functional assumptions on
the class of smoothers and completes construction of the test statistic.
Section 4 states basic distributional assumptions on parameter estimators
and presents asymptotic null distribution of the test statistic. Section 5
establishes key results on asymptotic size of the test. Section 6 studies
test consistency and local power. Section 7 presents results of some Monte
Carlo simulation studies. Section 8 concludes. Appendix A derives the
details of an adjustment component of the test statistic. Appendix B
provides proofs of theoretical results of the paper. Appendix C gives
examples of covariance matrix singularity and illustrates how they can fit
into our framework.

\section{Recentering, Selection and Smoothing in Inequality Tests \label%
{nor_s2}}

Let $\mu =(\mu _{1},\mu _{2},...,\mu _{p})^{\prime }$ be a column vector of
(functions of) parameters appearing in an econometric model. We are
interested in testing : 
\begin{equation}
H_{0}:\mu _{j}\geq 0\text{ for all }j\in \{1,2,...,p\}\text{ versus }%
H_{1}:\mu _{j}<0\text{ for at least one }j.  \label{t1}
\end{equation}%
We assume that there exists a vector $\widehat{\mu }$ of parameter
estimators based on sample size $T$ such that $\sqrt{T}(\widehat{\mu }-\mu )$
is asymptotically multivariate normal with mean 0 and covariance $V$
consistently estimated by $\widehat{V}$. The vector $\mu $ and matrix $V$
may depend on common parameters but this is generally kept implicit for
notational simplicity.

\subsection{Recentering and Generalized Moment Selection in Critical Value
Estimation \label{RL_1}}

Recent improved tests developed by Andrews and Soares (2010) of the
hypothesis (\ref{t1}) are distinguished by their use of estimated critical
values embodying a selection rule to statistically decide which inequalities
are binding ($\mu _{j}=0$). In brief, these tests proceed operationally as
follows. A statistic $S(\sqrt{T}\widehat{\mu },\widehat{V})$ is first
computed for some fixed function $S(.,.)$. The asymptotic critical value of
the statistic is then obtained by simulation (or resampling) as the
appropriate quantile of the distribution of $S(Z+K(T)\widetilde{\mu },%
\widehat{V})$ where $Z$ is an artificially generated vector such that $Z\sim
N(0,\widehat{V})$ conditionally on data, $\widetilde{\mu }$ is a recentered
null-compatible mean and $K(T)=o(\sqrt{T})$ is some positive
"tuning\textquotedblright\ function increasing without bound as $%
T\longrightarrow \infty $. Basic recentering defines $\widetilde{\mu }_{j}=0$
for $K(T)\widehat{\mu }_{j}\leq 1$. Setting $\widetilde{\mu }_{j}=0$ amounts
to selecting $j$ as the index of a binding constraint. For $K(T)\widehat{\mu 
}_{j}>1$, $\widetilde{\mu }_{j}$ is defined to ensure $K(T)$ $\widetilde{\mu 
}_{j}\longrightarrow \infty $ as $T\longrightarrow \infty $, this being
simply achieved by taking $\widetilde{\mu }_{j}=\widehat{\mu }$. Basic
selection as stated here is a special case of the Andrews and Soares (2010)
Generalized Moment Selection (GMS) procedure.\footnote{%
Indeed, this selection rule corresponds to use of moment selection function $%
\varphi _{j}^{(2)}$ considered by Andrews and Soares (2010, pp.\thinspace
131-132) with due allowance for standardization of parameter estimates. See
also Andrews and Barwick (2012, pp.\thinspace 8-9) for various examples of
the GMS selection rules.} \medskip

Data-dependent selection of binding constraints reduces possible
inefficiencies arising from fixing all the elements of $\widetilde{\mu }$ to
be zero (least favorable). On the other hand, regardless of how $\widetilde{%
\mu }$ is constructed, simulation (or bootstrap) is still needed since the
asymptotic distribution of the statistic used in this literature is
generally non-standard. This applies even to test statistics which aggregate
individual discrepancy values $\min (\widehat{\mu }_{j},0)$ in a simple
manner. They include the extreme value form studied by Hansen (2005) and the
sum 
\begin{equation}
\dsum\limits_{j=1}^{p}[-\sqrt{T}\min (\widehat{\mu }_{j},0)]  \label{L1}
\end{equation}%
lying within the very wide class of right-tailed tests studied by Andrews
and Soares (2010).

\subsection{The Smoothed Indicator Approach\label{smoothing}}

Let $1\{.\}$ denote the indicator taking value unity if the statement inside
the bracket is true and zero otherwise. The root cause of non-standard
distribution of (\ref{L1}) is the discontinuity at the origin of the
indicator $1\{x\leq 0\}$ underlying the negative-part function $\min
(x,0)=1\{x\leq 0\}x$. To overcome this problem, the present paper
investigates an indicator smoothing approach as follows. \medskip

First, we approximate the function $\min (x,0)$ by $\Psi _{T}(x)x$ where $%
\{\Psi _{T}(x)\}$ is a sequence of non-negative and non-increasing functions
each of which is continuously differentiable at the origin and converges
pointwise (except possibly at the origin) as $T\longrightarrow \infty $ \ to
the indicator function $1\{x\leq 0\}$. We refer to $\Psi _{T}(x)$ as an
(origin-smoothed) indicator smoother or a smoothed indicator for $1\{x\leq
0\}$. \medskip

In this paper, we will focus on the class of smoothed indicators generated
as $\Psi _{T}(x)=\Psi (K(T)x)$ for some fixed function $\Psi $ and a
\textquotedblleft tuner\textquotedblright\ $K(T)$ of the type mentioned in
Subsection \ref{RL_1}. The functional form of $\Psi $ includes decumulative
distribution functions for continuous variates as well as discrete yet
origin-smooth functions. We therefore replace the individual negative-part
statistic $\sqrt{T}\min (\widehat{\mu }_{j},0)$ of (\ref{L1}) by$\sqrt{T}%
\Psi _{T}(\widehat{\mu }_{j})\widehat{\mu }_{j}$. Subject to regularity
conditions set out later, $\Psi _{T}(\widehat{\mu }_{j})=o_{p}(1/\sqrt{T})$
for strictly positive $\mu _{j}$ and hence the term $\sqrt{T}\Psi _{T}(%
\widehat{\mu }_{j})\widehat{\mu }_{j}$ vanishes asymptotically. For
zero-valued $\mu _{j}$, $\Psi _{T}(\widehat{\mu }_{j})$ tends to $\Psi (0)$
in probability and $\sqrt{T}\Psi _{T}(\widehat{\mu }_{j})\widehat{\mu }_{j}$
is asymptotically equivalent to $\Psi (0)\sqrt{T}\widehat{\mu }_{j}$.
\medskip

Second, we consider a left-tailed test based on the statistic that replaces (%
\ref{L1}) with%
\begin{equation}
\dsum\limits_{j=1}^{p}\left[ \sqrt{T}\Psi _{T}(\widehat{\mu }_{j})\widehat{%
\mu }_{j}-\Lambda _{T}(\widehat{\mu }_{j},\widehat{v}_{jj})\right]
\label{pseudo local statistic}
\end{equation}%
where $\widehat{v}_{jj}$ is the $j$th diagonal element of $\widehat{V}$ and $%
\Lambda _{T}$ is an adjustment term approximating the expectation of $[\Psi
_{T}(\widehat{\mu }_{j})-\Psi (0)]\sqrt{T}\widehat{\mu }_{j}$ evaluated at $%
\mu _{j}=0$. This expectation is non-positive, though shrinking to zero in
large samples.\footnote{%
Note that $\Psi _{T}(\widehat{\mu }_{j})\widehat{\mu }_{j}\leq \Psi (0)%
\widehat{\mu }_{j}$ for any $T$ because the function $\Psi _{T}(x)=\Psi
(K(T)x)$ is constructed to be non-negative and non-increasing in $x$.} Under
suitable regularity conditions $\Lambda _{T}$, whose detailed construction
is given in Section \ref{smoothed indicator framework}, is non-positive for
all $T$ but converges to zero in probability. Hence, under the null
hypothesis the statistic (\ref{pseudo local statistic}) will be
asymptotically either degenerate or equivalent in distribution to a normal
variate and thus critical values for a test using (\ref{pseudo local
statistic}) will not require simulation. \medskip

Besides indicator smoothing, it is also appropriate to view $\Psi _{T}$ as a
form of binding inequality selection akin to the aforementioned GMS
procedure. The smoothed indicators in (\ref{pseudo local statistic})
essentially embed a data driven weighting scheme which automatically
concentrates the statistic (\ref{pseudo local statistic}) onto those
parameter estimates signaling binding inequalities. Indeed, consider the
specific smoothed indicator constructed as $\Psi _{T}(x)=1\{K(T)x\leq 1\}$.
Such $\Psi _{T}(x)$ simply shifts the point of discontinuity away from the
origin whilst still acting as a pure zero-one selector. Then the GMS based
recentering described in Subsection \ref{RL_1} would amount to setting $%
\widetilde{\mu }_{j}=(1-\Psi _{T}(\widehat{\mu }_{j}))\widehat{\mu }_{j}$.
In this case, the statistic (\ref{pseudo local statistic}) is equal to 
\begin{equation}
\dsum\limits_{j=1}^{p}\sqrt{T}(\widehat{\mu }_{j}-\widetilde{\mu }%
_{j})+o_{p}(1).  \label{smoothed test}
\end{equation}%
Since both $\widehat{\mu }$ and $\widetilde{\mu }$ are available as a
by-product of the mainstream tests of Subsection \ref{RL_1}, one may as well
perform a test on their difference. The asymptotic distribution of (\ref%
{smoothed test}) does not itself require simulation and recentering, so
there is no circularity of argument. Though (\ref{smoothed test}) and the
GMS test procedure are closely related, it is important to stress that the
present test enforces data driven selection of binding inequalities through
smoothed indicators within the test statistic itself rather than at the
stage of critical value estimation. Therefore, the class of statistics (\ref%
{pseudo local statistic}) does not lie in the otherwise very wide class
covered by the work of Andrews and Soares (2010). \medskip

It is worth noting that the approach to achieve asymptotic normality in this
paper is distinct from alternative devices such as those of Dykstra (1991)
and Menzel (2008) who demonstrate that even the $QLR$ statistic can be
asymptotically normal when $p$, the dimension of $\mu $, is viewed as
increasing with $T$ to infinity. Recent papers by Lee and Whang (2009) and
Lee, Song and Whang (2011) obtain asymptotic normality for a class of
functional inequality test statistics. Their particular device
(poissonization) requires $\mu $ to be infinitely dimensional at the outset.
By contrast, in the framework of testing finite and fixed $p$ inequalities,
the present paper (and its preliminary versions (Chen and Szroeter (2006,
2009) and Chen (2009, Chapter 3)) where a prototype asymptotically normal
test statistic appears) uses only large $T$ asymptotics and an indicator
smoothing device. The strategy adopted by this work in testing is akin to
Horowitz (1992) who sought to resolve non-standard asymptotic behavior in
estimation by replacing a discrete indicator function with a smoothed
version. Therefore, the smoothing mechanism investigated by this paper to
obtain standard asymptotic distribution results could also be of theoretical
interest in its own right.

\section{Smoothed Indicator Class and Test Procedure\label{smoothed
indicator framework}}

We now formally set out regularity conditions on the smoothed indicator $%
\Psi _{T}(x)$, $x\in R$. We require that 
\begin{equation}
\Psi _{T}(x)=\Psi (K(T)x)  \label{psiT}
\end{equation}%
where $\Psi (.)$ and $K(T)$ are functions satisfying the following
assumptions:

\bigskip

[A1] \ \ \ $\Psi (x)$\textit{\ is a non-increasing function and }$0\leq \Psi
(x)\leq 1$\textit{\ for }$x\in R\medskip .$

[A2] \ \ \ $\Psi (0)>0$\textit{\ and, throughout some open interval
containing }$x=0$\textit{\ and at all except}

\textit{\ \ \ \ \ \ \ \ \ \ possibly\ a finite number of points outside
that\ interval,}$\ \Psi (x)$\textit{\ has a continuous}

\textit{\ \ \ \ \ \ \ \ \ \ first derivative }$\psi (x)$\textit{\ that is
bounded absolutely by a finite\ positive\ constant.}

\textit{\ \ \ \ \ \ \ \ \ \ The left-hand limits of }$\psi (y)$\textit{\ as }%
$y$\textit{\ approaches }$x$\textit{\ exist at any }$x\in R$\textit{.}$%
\medskip $

[A3] \ \ \ $K(T)$\textit{\ is positive and increasing in }$T\medskip .$

[A4] \ \ \ $K(T)\longrightarrow \infty $\textit{\ and \ }$K(T)/\sqrt{T}%
\longrightarrow 0$\textit{\ as }$T\longrightarrow \infty \medskip .$

[A5] \ \ \ $\Psi (x)\longrightarrow 1$\textit{\ as }$x\longrightarrow
-\infty \medskip .$

[A6] \ \ \ $\sqrt{T}\Psi (K(T)x)\longrightarrow 0$\textit{\ as }$%
T\longrightarrow \infty $\textit{\ for }$x>0\medskip .$

\bigskip

Assumptions [A1]-[A6] are very mild and satisfied by all the particular $%
\Psi $ functions including step-at-unity, logistic and normal, discussed in
Section \ref{nor_s5} and used in the simulations of this paper. Assumption
[A4] regulates the rate at which the \textquotedblleft
tuning\textquotedblright\ parameter $K(T)$ can grow and, in the context of
Andrews and Soares (2010) discussed in Subsection \ref{RL_1}, enables
consistent selection of binding constraints. Forms of tuning are also used
by Chernozhukov et al.\thinspace (2007) and Linton et al.\thinspace (2010).
[A2] enables smoothing for asymptotic normality through zero-valued $\mu
_{j} $, whilst [A6] creates data-driven importance weighting in the sense
that each $\widehat{\mu }_{j}$ corresponding to strictly positive $\mu _{j}$
is likely to contribute ever less to the value of the test statistic as $T$
increases. In consequence, the statistic will be asymptotically dominated by
those $\widehat{\mu }_{j}$ corresponding to zero or negative $\mu _{j}$,
detection of which is the very purpose of the test. \medskip

To implement the test, we have to construct the term $\Lambda _{T}$ in (\ref%
{pseudo local statistic}) of Subsection \ref{smoothing}. Though Assumptions
[A2], [A4] and (\ref{psiT}) above are given so that, for $\mu _{j}=0$, $%
\sqrt{T}\Psi _{T}(\widehat{\mu }_{j})\widehat{\mu }_{j}$ in (\ref{pseudo
local statistic}) is asymptotically equivalent to $\Psi (0)\sqrt{T}\widehat{%
\mu }_{j}$, the difference $\sqrt{T}\Psi _{T}(\widehat{\mu }_{j})\widehat{%
\mu }_{j}-\Psi (0)\sqrt{T}\widehat{\mu }_{j}$ remains non-positive in large
samples. Whilst asymptotically negligible, this may be size-distorting in
finite samples. To systematically offset that effect, the adjustment term $%
\Lambda _{T}$ is constructed as follows to approximate the expectation of $%
[\Psi _{T}(\widehat{\mu }_{j})-\Psi (0)]\sqrt{T}\widehat{\mu }_{j}$.
\medskip\ 

Under Assumption [A2], there are finite increasing values $a_{1},...,a_{n}$
for some $n\geq 1$ such that $\Psi (x)$ is continuously differentiable in
intervals $(-\infty ,a_{1}),(a_{1},a_{2}),...,(a_{n},\infty )$. Because $%
\Psi $ is bounded and non-increasing, its one-sided limits $\Psi
(a_{i}^{-})\equiv \lim_{x\longrightarrow a_{i}^{-}}\Psi (x)$ and $\Psi
(a_{i}^{+})\equiv \lim_{x\longrightarrow a_{i}^{+}}\Psi (x)$ for $i\in
\{1,2,...,n\}$ exist. Let $\widetilde{\psi }(x)$, $x\in R$ be the "extended"
derivative of $\Psi $ defined as the left-hand limit of $\psi (x)$. Namely, $%
\widetilde{\psi }(x)\equiv \lim_{y\longrightarrow x^{-}}\psi (y)$. Then the
algebraic form of $\Lambda _{T}$ whose detailed derivation is given in
Appendix A can be written as%
\begin{equation}
\Lambda _{T}(\widehat{\mu }_{j},\widehat{v}_{jj})=\widehat{v}_{jj}\widetilde{%
\psi }(K(T)\widehat{\mu }_{j})K(T)/\sqrt{T}-\sqrt{\widehat{v}_{jj}}%
\sum_{i=1}^{n}(\Psi (a_{i}^{-})-\Psi (a_{i}^{+}))\phi (\frac{a_{i}\sqrt{T}}{%
\sqrt{\widehat{v}_{jj}}K(T)})  \label{gammaT}
\end{equation}%
where $\phi $ is the standard normal density function.\medskip

For the simple choice $\Psi (x)=1\{x\leq 1\}$ used to form the statistic (%
\ref{smoothed test}), $\widetilde{\psi }=0$ and there is a single
discontinuity at $x=1$ so the proxy simplifies to 
\begin{equation}
\Lambda _{T}(\widehat{\mu }_{j},\widehat{v}_{jj})=-\sqrt{\widehat{v}_{jj}}%
\phi (\frac{\sqrt{T}}{\sqrt{\widehat{v}_{jj}}K(T)}).  \label{g1}
\end{equation}%
On the other hand, for everywhere continuously differentiable $\Psi $, $%
\widetilde{\psi }(x)=\psi (x)$ for $x\in R$ and $\Psi (a_{i}^{-})=\Psi
(a_{i}^{+})$ for $i\in \{1,2,...,n\}.$ Hence $\Lambda _{T}$ for such case
simplifies to 
\begin{equation}
\Lambda _{T}(\widehat{\mu }_{j},\widehat{v}_{jj})=\widehat{v}_{jj}\psi (K(T)%
\widehat{\mu }_{j})K(T)/\sqrt{T}\text{.}  \label{g2}
\end{equation}%
Note that since $\Psi $ is non-increasing, for any $T$, $\Lambda _{T}(%
\widehat{\mu }_{j},\widehat{v}_{jj})$ given by (\ref{gammaT}) is
non-positive by construction. Besides, under Assumption [A4] $\Lambda _{T}(%
\widehat{\mu }_{j},\widehat{v}_{jj})$ tends to zero in probability as $T$
tends to infinity. Hence for those $\mu _{j}\neq 0$, the impact of adjusting 
$\sqrt{T}\Psi _{T}(\widehat{\mu }_{j})\widehat{\mu }_{j}$ with the term $%
\Lambda _{T}(\widehat{\mu }_{j},\widehat{v}_{jj})$ on test behavior is
asymptotically negligible though the adjustment (\ref{gammaT}) is applied
for each $j\in \{1,2,..,p\}$. \medskip

Finally, we consider a further useful generalization by replacing each $%
\widehat{\mu }_{j}$ in (\ref{pseudo local statistic}) with $\widehat{\theta }%
_{j}\widehat{\mu }_{j}$ for any positive scalar $\widehat{\theta }_{j}$,
which can be fixed known or estimated. Choosing $\widehat{\theta }_{j}$ to
be inverse of the estimated asymptotic standard deviation of $\widehat{\mu }%
_{j}$ amounts to conducting the test on t-ratios. Other choices of $\widehat{%
\theta }_{j}$ are discussed in Appendix C which deals with estimator
covariance singularity issues. With this enhancing feature, the adjustment
term $\Lambda _{T}(\widehat{\mu }_{j},\widehat{v}_{jj})$ is replaced by $%
\Lambda _{T}(\widehat{\theta }_{j}\widehat{\mu }_{j},\widehat{\theta }%
_{j}^{2}\widehat{v}_{jj})$. We now present the test procedure as
follows.\medskip

Let $\widehat{\Psi },\widehat{\Lambda },e_{p}$ be the $p$ dimensional column
vectors and $\widehat{\Delta }$ be the diagonal matrix defined as\ 
\begin{eqnarray}
\widehat{\Psi } &\equiv &(\Psi (K(T)\widehat{\theta }_{1}\widehat{\mu }%
_{1}),\Psi (K(T)\widehat{\theta }_{2}\widehat{\mu }_{2}),...,\Psi (K(T)%
\widehat{\theta }_{p}\widehat{\mu }_{p}))^{\prime },  \label{psi_hat} \\
\widehat{\Lambda } &\equiv &(\Lambda _{T}(\widehat{\theta }_{1}\widehat{\mu }%
_{1},\widehat{\theta }_{1}^{2}\widehat{v}_{11}),\Lambda _{T}(\widehat{\theta 
}_{2}\widehat{\mu }_{2},\widehat{\theta }_{2}^{2}\widehat{v}%
_{22}),...,\Lambda _{T}(\widehat{\theta }_{p}\widehat{\mu }_{p},\widehat{%
\theta }_{p}^{2}\widehat{v}_{pp}))^{\prime },  \label{gamma_hat} \\
e_{p} &\equiv &(1,1,...,1)^{\prime }, \\
\text{\ }\widehat{\Delta } &\equiv &diag(\widehat{\theta }_{1},\widehat{%
\theta }_{2},...,\widehat{\theta }_{p}).
\end{eqnarray}

Let 
\begin{eqnarray}
Q_{1} &\equiv &\sqrt{T}\widehat{\Psi }^{\prime }\widehat{\Delta }\widehat{%
\mu }-e_{p}^{\prime }\widehat{\Lambda }  \label{q1} \\
Q_{2} &\equiv &\sqrt{\widehat{\Psi }^{\prime }\widehat{\Delta }\widehat{V}%
\widehat{\Delta }\widehat{\Psi }}.  \label{q2}
\end{eqnarray}%
We define the test statistic as \smallskip 
\begin{equation}
Q=\left\{ 
\begin{array}{c}
\Phi (Q_{1}/Q_{2})\text{ \ \ if }Q_{2}>0 \\ 
1\text{ \ \ \ \ \ \ \ \ \ \ \ \ \ \ \ if }Q_{2}=0%
\end{array}%
\right.  \label{q}
\end{equation}%
\medskip where $\Phi (x)$ is the standard normal distribution function. For
asymptotic significance level $\alpha $, we reject $H_{0}$ if $\ Q<\alpha $.
The test statistic $Q$ is therefore a form of tail probability or
p-value.\medskip

We now sketch the reasoning which validates the test. Formal theorems are
given later. Intuitively, we should reject $H_{0}$ if $Q_{1}$ is too small.
For those parameter points under $H_{0}$ for which the probability limit of $%
Q_{2}$ is nonzero, $Q_{2}$ will be strictly positive with probability
approaching one. Then the ratio $Q_{1}/Q_{2}$ will exist and be
asymptotically normal. By contrast, for all points under $H_{1}$, the value
of $Q_{1}$ will go in probability to minus infinity. Therefore, in cases
where $Q_{2}$ is positive, we propose to reject $H_{0}$ if $Q_{1}/Q_{2}$ is
too small compared with the normal distribution.\medskip

Note that our assumptions on the smoothed indicators do not rule out
discrete but origin-smooth $\Psi $ functions such as the step-at-unity
example of Section \ref{nor_s5}. For such a discrete function, $\widehat{%
\Psi }$ will be a null vector with probability approaching one when all $\mu
_{j},$ $j\in \{1,2,...,p\},$ are strictly positive. In this case, $Q_{2}$ is
also zero by (\ref{q2}) with probability approaching one. Therefore,
occurrence of the event $Q_{2}=0$ is possible and signals that we should not
reject $H_{0}$. Note that it is not an adhoc choice to set $Q=1$ when $%
Q_{2}=0$ occurs because the probability limit of $\Phi (Q_{1}/Q_{2})$ is
also one when all $\mu _{j}^{{}}$ parameters are strictly positive and $\Psi 
$ is an everywhere positive function.\footnote{%
The case of $\Psi $ being everywhere positive is more complicated because $%
Q_{2}$ can then be almost surely strictly positive. If all $\mu _{j}^{{}}$
parameters are strictly positive, both numerator and denominator in the
ratio $Q_{1}/Q_{2}$ tend to zero in probability. See Appendix B.4 for
analysis of the asymptotic properties of the test statistic $Q$ in that case.%
}

\section{Distributional Assumptions and Asymptotic Null Distribution\label%
{nor_s3}}

We begin by stating the following high-level assumptions which enable us to
derive some basic asymptotic properties of the test. Except for [D2], these
assumptions are standard.\medskip\ 

Define $\Delta $ as the diagonal matrix $\Delta \equiv diag(\theta
_{1},\theta _{2},...,\theta _{p})$ where $\theta _{j}$ is strictly positive
and its estimator $\widehat{\theta }_{j}$ is almost surely strictly positive
for $j\in \{1,2,...,p\}$. Let $d(\mu )$ be defined as the $p$ dimensional
vector whose $j$th element equals 0, $\Psi (0)$, 1 when $\mu _{j}>0$, $\mu
_{j}=0$, $\mu _{j}<0$ respectively. For notational simplicity, we keep
implicit the possible dependence of the true values of the parameters $\mu $%
, $V$ and $\Delta $ on the underlying data generating process. \medskip

We assume that, as $T$ tends to infinity,\bigskip

[D1] $\ \ \ \ \ \ \ \ \sqrt{T}(\widehat{\mu }-\mu )\overset{d}{%
\longrightarrow }N(0,V)$\textit{\ where }$V$\textit{\ is some finite
positive semi-definite matrix.}$\bigskip $ \newline
The variance $V$ need not be invertible but must satisfy the following
condition (whose verification is illustrated in Appendix C).\bigskip

[D2] \ \ \ \ \ \ \ $V\Delta d(\mu )\neq 0$\textit{\ \ for non-zero }$d(\mu )$%
\textit{.}$\bigskip $\newline
Assumption [D2] amounts to saying that the asymptotic distribution of$\sqrt{T%
}d(\mu )^{\prime }\Delta (\widehat{\mu }-\mu )$ should not be degenerate. $%
\bigskip $

[D3] \ \ \ \ \ \ \ \ $\widehat{V}\overset{p}{\longrightarrow }V$\textit{\
for some almost surely positive semi-definite estimator }$\widehat{V}$%
\textit{.}$\bigskip $

[D4] \ \ \ \ \ \ \ $\ \widehat{\Delta }\overset{p}{\longrightarrow }\Delta
.\bigskip $

Now let $J$ denote the set $\{1,2,...,p\}$ and decompose this as $J=A\cup
M\cup B$, where 
\begin{equation*}
A\equiv \{j\in J:\mu _{j}^{{}}>0\},\text{ }M\equiv \{j\in J:\mu
_{j}^{{}}=0\},\text{ }B\equiv \{j\in J:\mu _{j}^{{}}<0\}.
\end{equation*}%
Let $U(0,1)$ denote a scalar random variable that is uniformly distributed
in the interval $[0,1]$. We now present the asymptotic null distribution of
the test statistic.

\begin{theorem}[Pointwise Asymptotic Null Distribution]
\label{ch_nor_thm1}\textit{Given [A1], [A2], [A3], [A4], [A6] with [D1] -
[D4], the following are true under }$H_{0}:\mu _{j}\geq 0$\textit{\ for all }%
$j\in J$\textit{\ with limits taken along }$T\longrightarrow \infty .$%
\newline
\qquad \qquad \qquad \qquad \qquad (1)\textit{\ \ If }$M\neq \varnothing $%
\textit{, then }$Q$\textit{\ }$\overset{d}{\longrightarrow }$\textit{\ }$%
U(0,1).\smallskip $\newline
\qquad \qquad \qquad \qquad \qquad (2)\textit{\ \ If }$M=\varnothing $%
\textit{, then }$Q$\textit{\ }$\overset{p}{\longrightarrow }$\textit{\ }$1.$
\end{theorem}

Part (1) of this theorem reflects the fact that, for any fixed data
generating process whose $\mu $ value lies on the boundary of null
hypothesis space, the distribution of the test statistic $Q$ is
asymptotically non-degenerate and given (\ref{q}), the limiting distribution
of the ratio $Q_{1}/Q_{2}$ is standard normal. This justifies the idea of
smoothing for normality. Moreover, $Q$ has the same limiting distribution at
each boundary point. Part (2) says that, at any fixed data generating
process whose $\mu $ value lies in the interior of null hypothesis space,
the asymptotic distribution of $Q$ is degenerate and $Q$ will take value
above $\alpha $ with probability tending to 1.

\section{Asymptotic Test Size \label{nor_s4}}

\subsection{Pointwise and Uniform Asymptotic Control of Test Size}

Theorem \ref{ch_nor_thm1} shows that the test statistic $Q$ is not
asymptotically pivotal since its limiting distribution and hence the
asymptotic null rejection probability depend on the true value of $\mu $. By
definition, the \textit{pointwise} asymptotic size of the test is the
supremum of the asymptotic rejection probability viewed as a function of $%
\mu $ on the domain defined by $H_{0}$. So Theorem \ref{ch_nor_thm1} implies
that this size equals the nominal level $\alpha $ and hence the test is
asymptotically exact in the pointwise sense. However, pointwise asymptotic
exactness is a weak property. It is desirable to ensure the convergence of
the test size to the nominal level holds uniformly over the null-restricted
parameter and data distribution spaces. In this section we present results
showing that the test size is asymptotically exact in the uniform
sense.\medskip

To distinguish between pointwise and uniform modes of analysis, we need some
additional notation. Note that parameters such as $\mu $ and $V$ are
functionals of the underlying data generating distribution. Suppose the data
consist of i.i.d.\thinspace vectors $x_{t}$ ($t=1,...,T$) drawn from a joint
distribution $G$. We henceforth use the notation $P_{G}(.)$ to make explicit
the dependence of probability on $G$. Let $\Gamma $ denote the set of all
possible $G$ compatible with prior knowledge or presumed specification of
the data generating process. Then Assumptions [D1] - [D4] amount to
restrictions characterizing the class $\Gamma $. Let $\Gamma _{0}$ be the
subset of $\Gamma $ that satisfies the null hypothesis. In the present test
procedure, "$Q<\alpha $\textquotedblright\ is synonymous with
\textquotedblleft $Q$ rejects $H_{0}$\textquotedblright . Hence, the
rejection probability of the test is $P_{G}(Q<\alpha )$ and the finite
sample test size is $\sup_{G\in \Gamma _{0}}P_{G}(Q<\alpha )$. \medskip

Though Theorem \ref{ch_nor_thm1} implies that convergence of rejection
probability is not uniform over $G\in \Gamma _{0}$, the test can be shown to
be uniformly asymptotically level $\alpha $ (Lehmann and Romano (2005, p.
422)) in the sense that 
\begin{equation}
\limsup_{T\longrightarrow \infty }\sup_{G\in \Gamma _{0}}P_{G}(Q<\alpha
)\leq \alpha .  \label{c2}
\end{equation}%
Inequality (\ref{c2}) and Part (1) of Theorem \ref{ch_nor_thm1} together
imply the test size is asymptotically exact in the uniform sense that%
\begin{equation}
\limsup_{T\longrightarrow \infty }\sup_{G\in \Gamma _{0}}P_{G}(Q<\alpha
)=\alpha .  \label{c3}
\end{equation}%
The property (\ref{c3}) is important for the asymptotic size to be a good
approximation to the finite-sample size of the test.\footnote{%
Note that the notion of asymptotic test size using $\limsup_{T%
\longrightarrow \infty }\sup_{G\in \Gamma _{0}}P_{G}(Q<\alpha )$ is stronger
than its pointwise version $\sup_{G\in \Gamma _{0}}\limsup_{T\longrightarrow
\infty }P_{G}(Q<\alpha ).$ See Lehmann and Romano (2005, p. 422) for an
illustrating example in which pointwise asymptotic size can be a very poor
approximation to the finite sample test size.} Such uniformity property has
been emphasized in recent literature (e.g., see Mikusheva (2007), Romano and
Shaikh (2008), Andrews and Guggenberger (2009) and Andrews and Soares
(2010)) particularly when limit behavior of the test statistic can be
discontinuous. Accordingly, we establish the validity of (\ref{c3}) in
Theorem \ref{uniform size convergence}. \medskip

Before presenting the formal regularity conditions ensuring (\ref{c3}), we
explain here how (\ref{c3}) is possible despite asymptotic non-pivotality of
the test statistic. First note that by (\ref{q}), 
\begin{equation}
P_{G}(Q<\alpha )\leq P_{G}(Q_{1}-z_{\alpha }Q_{2}<0)  \label{c}
\end{equation}%
where $z_{\alpha }$\ is the $\alpha $\ quantile of the standard normal
distribution\textit{.} The transformed statistic $(Q_{1}-z_{\alpha }Q_{2})$
is still not asymptotically pivotal but it can be shown that, given any
arbitrary sufficiently small (relative to model constants) positive scalar $%
\eta $, we have with probability at least $\left( 1-\eta \right) $ for all
sufficiently large $T$ that 
\begin{equation*}
Q_{1}-z_{\alpha }Q_{2}\geq r_{T}^{\prime }\sqrt{T}(\widehat{\mu }-\mu
)-(z_{\alpha }c_{2}(\eta )+c_{1}(\eta ))\sqrt{r_{T}^{\prime }Vr_{T}}
\end{equation*}%
where $r_{T}$, $\mu $ and $V$ are non-stochastic $G$-dependent quantities
such that either $r_{T}=0$ or $r_{T}^{\prime }Vr_{T}$ is bounded away from
zero over $G\in \Gamma _{0}$, whilst $c_{1}(\eta )$ and $c_{2}(\eta )$ are
non-stochastic functions that do not depend on $G$ and $c_{1}(\eta
)\longrightarrow 0$ and $c_{2}(\eta )\longrightarrow 1$ as $\eta
\longrightarrow 0$. Therefore,%
\begin{equation}
P_{G}(Q_{1}-z_{\alpha }Q_{2}<0)\leq P_{G}(r_{T}^{\prime }\sqrt{T}(\widehat{%
\mu }-\mu )<(z_{\alpha }c_{2}(\eta )+c_{1}(\eta ))\sqrt{r_{T}^{\prime }Vr_{T}%
})+\eta  \label{c4}
\end{equation}%
whose right hand will tend, uniformly over $G$ giving non-zero $r_{T}$, to $%
\Phi (z_{\alpha }c_{2}(\eta )+c_{1}(\eta ))+\eta $ which is also
automatically a weak upper bound on (\ref{c4}) for the case $r_{T}=0$. This
uniformly valid probability bound therefore applies to (\ref{c}) for
arbitrarily small $\eta $ hence implies that (\ref{c2}) holds. Equality is
obtained by invoking Theorem \ref{ch_nor_thm1} which says $\alpha $ is
actually attained as the limit of $P_{G}(Q<\alpha )$ evaluated at any fixed $%
G\in \Gamma _{0}$ whose $\mu $ has at least a zero-valued element. \medskip

The explanation provided above is indicative but short of a formal proof. In
the next subsection we present additional \textquotedblleft
uniform\textquotedblright\ assumptions, strengthening the existing
\textquotedblleft pointwise\textquotedblright\ assumptions [D1] - [D4] of
Section \ref{nor_s3}, that are needed to make the argument rigorous. The
full proof, along with examples to illustrate some of the assumptions, will
be found in the Appendix B.

\subsection{Uniform Asymptotic Exactness of Test Size}

In this section we rigorously address the issue of asymptotic exactness of
test size in the uniform sense given by (\ref{c3}). For this purpose, we
strengthen Assumptions [D1] - [D4] by the following Assumptions [U1] - [U4]
where objects such as $K(T)$ have already been defined in Assumptions [A1] -
[A6]. Define the vector $Y$ and the scalar $\delta _{T}$ as

\begin{equation*}
Y\equiv \sqrt{T}(\widehat{\mu }-\mu ),\text{ \ \ \ \ }\delta _{T}\equiv 
\sqrt{K(T)/\sqrt{T}}.
\end{equation*}%
Note that Assumption [A4] implies that \ $\delta _{T}\longrightarrow 0$ as $%
T\longrightarrow \infty $. For any matrix $m$, let $\left\Vert m\right\Vert
\equiv \max \{\left\vert m_{ij}\right\vert \}$ where $m_{ij}$\textit{\ }%
denotes the $(i,j)$-th element of $m.$

\bigskip

\textbf{Assumption} [U1] : \textit{For any finite scalar value }$\eta >0$%
\textit{,} 
\begin{equation*}
\lim_{T\longrightarrow \infty }\inf_{G\in \Gamma _{0}}P_{G}(\delta
_{T}\left\Vert Y\right\Vert <\eta ,\text{ }||\widehat{V}-V_{G}||\text{ }%
<\eta )=1.
\end{equation*}

\bigskip

\textbf{Assumption} [U2] : \textit{Let }$\Phi (.)$\textit{\ denote the
standard normal distribution function. Then given any finite scalar }$c$%
\textit{,}%
\begin{equation}
\lim_{T\longrightarrow \infty }\sup_{G\in \Gamma _{0}}\sup_{\beta :\beta
^{\prime }V_{G}\beta =1}|P_{G}(\beta ^{\prime }Y\leq c)-\Phi (c)|\text{ }=0.
\label{U3}
\end{equation}

\bigskip

To illustrate how the high-level Assumptions [U1] and [U2] may be verified,
consider the leading example where $\widehat{\mu }$ and $\widehat{V}$ are
the sample mean and variance of i.i.d.\thinspace random vectors $x_{t}$, $%
(t=1,2,...,T)$ with joint distribution $G$.\footnote{%
This simple average framework is used extensively in recent literature on
inference for (unconditional) moment inequality models. See, e.g.
Chernozhukov et al.\thinspace (2007), Romano and Shaikh (2008), Rosen
(2008), Andrews and Guggenberger (2009), Andrews and Soares (2010), Andrews
and Barwick (2012) and references cited therein.} Then the simple but not
necessarily the weakest primitive condition guaranteeing both Assumptions
[U1] and [U2] is that the first four moments of every element of $x_{t}$
exist and are bounded uniformly over $G\in \Gamma _{0}$. This condition
allows the application of the Chebychev inequality to components of the
right-hand side of the inequality%
\begin{equation*}
P_{G}(\delta _{T}\left\Vert Y\right\Vert <\eta ,\text{ }||\widehat{V}-V_{G}||%
\text{ }<\eta )\geq P_{G}(\delta _{T}\left\Vert Y\right\Vert <\eta )+P_{G}(||%
\widehat{V}-V_{G}||\text{ }<\eta )-1
\end{equation*}%
to deduce that Assumption [U1] holds. To verify Assumption [U2] we first
note that, by Lemma 4 proved in the Appendix, it is sufficient for (\ref{U3}%
) that 
\begin{equation}
\lim_{T\longrightarrow \infty }|P_{G_{T}}(\beta _{T}^{\prime }Y\leq c)-\Phi
(c)|\text{ }=0  \label{U}
\end{equation}%
for all non-stochastic sequences $(G_{T},\beta _{T})$ satisfying $G_{T}\in
\Gamma _{0}$ and $\beta _{T}^{\prime }V_{G_{T}}\beta _{T}=1$. By the i.i.d.
assumption, $\beta _{T}^{\prime }Y$ is $1/\sqrt{T}$ times the sum of $T$
variates $\beta _{T}^{\prime }(x_{t}-E_{G_{T}}(x_{t}))$ which are mutually
i.i.d. with mean 0 and variance 1 for each $T$ when $\beta _{T}^{\prime
}V_{G_{T}}\beta _{T}=1$. This meets the requirements of the double array
version of the classic Lindeberg-Feller central limit theorem thus
establishing asymptotic unit normality of $\beta _{T}^{\prime }Y$ hence
verifying (\ref{U}).\medskip

For the next assumption, recall that $\theta _{j}$ is the $j$th diagonal
element of the matrix $\Delta $. For notational simplicity, the general
dependence of $\theta _{j}$ and $\Delta $ on $G$ will be kept implicit.

\bigskip

\textbf{Assumption} [U3] : \textit{(i) There are finite positive scalars }$%
\lambda $\textit{\ and }$\lambda ^{\prime }$\textit{\ such that }$\lambda
^{\prime }\leq $\textit{\ }$\theta _{j}\leq \lambda ,$\textit{\ }$%
(j=1,2,...,p)$\textit{\ uniformly over }$G\in \Gamma _{0}$\textit{. (ii) For
any finite scalar value }$\eta >0$\textit{,}

\begin{equation*}
\lim_{T\longrightarrow \infty }\inf_{G\in \Gamma _{0}}P_{G}(\left\Vert 
\widehat{\Delta }-\Delta \right\Vert <\eta \delta _{T})=1.
\end{equation*}

\bigskip

Assumption [U3] holds automatically when $\Delta $ is numerically specified
by the user hence $\widehat{\Delta }=\Delta $. It also allows $\theta _{j}$
to be $1/\sqrt{v_{jj}}$ where $v_{jj}$ is the $j$th diagonal element of $%
V_{G}$ provided that $v_{jj}$ is bounded below by some constant, say $L>0$,
uniformly over $G\in \Gamma _{0}$.\footnote{%
Assumption [U3]-(ii) is stronger than requiring consistency of $\widehat{%
\theta }_{j}$ as an estimator of $\theta _{j}$. An alternative approach is
to strengthen Assumption [U2] by taking $Y$ to be $\sqrt{T}(\widehat{\Delta }%
\widehat{\mu }-\Delta \mu )$ rather than just $\sqrt{T}(\widehat{\mu }-\mu )$%
. But that would be implicitly assuming $\sqrt{T}(\widehat{\theta }%
_{j}-\theta _{j})$ is asymptotically normal (or degenerate). Such an
assumption is even stronger than [U3]-(ii) and quite unnecessary for our
results.} In such case, 
\begin{equation}
\left\vert \widehat{\theta }_{j}-\theta _{j}\right\vert \leq \left\vert 
\widehat{v}_{jj}-v_{jj}\right\vert \sqrt{2}L^{-3/2}  \label{U3 v}
\end{equation}%
when $\left\vert \widehat{v}_{jj}-v_{jj}\right\vert <L/2$.\footnote{%
By mean value expansion, $\left\vert \widehat{\theta }_{j}-\theta
_{j}\right\vert =\left\vert \widehat{v}_{jj}-v_{jj}\right\vert /(2|\overline{%
v}_{jj}|^{3/2})$ where $\overline{v}_{jj}$ lies between $\widehat{v}_{jj}$
and $v_{jj}$. Thus when $\left\vert \widehat{v}_{jj}-v_{jj}\right\vert <L/2$%
, inequality (\ref{U3 v}) follows by noting that $\left\vert \overline{v}%
_{jj}-v_{jj}\right\vert \leq \left\vert \widehat{v}_{jj}-v_{jj}\right\vert $.%
} Hence in the sample mean example described after Assumption [U2], we can
verify [U3]-(ii) by applying the Chebychev inequality to show that $%
P_{G}(\left\vert \widehat{v}_{jj}-v_{jj}\right\vert $ $<\eta \delta _{T})$
also tends to 1 uniformly over $G\in \Gamma _{0}$.\medskip

For any given positive scalar $\sigma $, let $d_{\sigma }(\mu )$ denote the $%
p$ dimensional vector whose $j$th element equals $\Psi (0)$ when $0\leq \mu
_{j}\leq \sigma $ and equals 0 otherwise.

\bigskip

\textbf{Assumption} [U4] : \textit{There are finite positive real scalars }$%
\omega $\textit{, }$\omega ^{\prime }$\textit{\ and }$\sigma $\textit{\ such
that the following hold uniformly over }$G\in \Gamma _{0}:$\textit{\ (i) }$%
\left\Vert V_{G}\right\Vert <\omega .$\textit{\ (ii) }$d_{\sigma }(\mu
)^{\prime }\Delta V_{G}\Delta d_{\sigma }(\mu )>\omega ^{\prime }$\textit{\
for all non-zero }$d_{\sigma }(\mu )$\textit{.}

\bigskip

Assumption [U4]-(i) is simply a boundedness assumption which automatically
holds when $V_{G}$ is a correlation matrix. [U4]-(ii) holds automatically
when the smallest eigenvalue of $V_{G}$ is bounded away from zero over $G\in
\Gamma _{0}$. Note that [U4]-(ii), essentially strengthening Assumption
[D2], requires that the asymptotic variance of $\sqrt{T}d_{\sigma }(\mu
)^{\prime }\Delta (\widehat{\mu }-\mu )$ be bounded away from zero for all
non-zero $d_{\sigma }(\mu )$. This is a high level assumption whose
verification will be illustrated in examples of Appendix C.\medskip

We can now present the following theorem establishing asymptotic exactness
of the test in the uniform sense.

\begin{theorem}[Uniform Asymptotic Exactness of Test Size]
\label{uniform size convergence}\textit{Given Assumptions [D1] - [D4],
suppose Assumptions [U1] - [U4] also hold. Assume some }$G\in \Gamma _{0}$%
\textit{\ has }$\mu $\textit{\ value containing at least one zero-valued
element. Then under Assumptions [A1], [A2], [A3], [A4], [A6] and given} $%
0<\alpha <1/2,$%
\begin{equation*}
\limsup_{T\longrightarrow \infty }\sup_{G\in \Gamma _{0}}P_{G}(Q<\alpha
)=\alpha .
\end{equation*}
\end{theorem}

\section{Asymptotic Power of the Test\label{power of the test}}

In this section, we study the asymptotic power properties of the test. Proof
of all results are presented in the Appendix. For notational simplicity, we
suppress the dependence of probability and parameters on the underlying data
generating distribution. We first show that the test is consistent against
fixed alternative hypotheses.

\begin{theorem}[Consistency]
\label{consistency theorem}\textit{Given [A1] - [A6] with [D1] - [D4], \ the
following is true under }$H_{1}:\mu _{j}^{{}}<0$\textit{\ for some }$j\in
\{1,2,...,p\}$.%
\begin{equation*}
P(Q<\alpha )\longrightarrow 1\text{ \ as \ }T\longrightarrow \infty \medskip
.
\end{equation*}
\end{theorem}

Besides consistency, we are also interested in the local behavior of the
test. In order to derive a local power function, we consider a sequence of $%
\mu $ values in the alternative-hypothesis space tending at rate $T^{-1/2}$
to a value $\gamma \equiv (\gamma _{1},\gamma _{2},...,\gamma _{p})^{\prime
} $ on the boundary of the null-hypothesis space. Specifically, we represent
the $j$th element of $\mu $ of such a local sequence as 
\begin{equation}
\mu _{j}^{{}}=\gamma _{j}+\frac{c_{j}}{\sqrt{T}}  \label{local}
\end{equation}%
where $\gamma _{j}\geq 0$ and $c_{j}$ are constants such that $\gamma _{j}=0$
and $c_{j}<0$ hold simultaneously for at least one $j$. The sequence (\ref%
{local}) is said to be \textit{core} if $c_{j}<0$ holds in every instance of 
$\gamma _{j}=0$. A core local sequence corresponds to Neyman-Pitman drift in
the original sense (McManus (1991)) whereby parameter values conflicting
with the null hypothesis are imagined \textit{ceteris paribus} to draw ever
closer to compliance as $T$ increases. In the easily-visualized case $p=2$,
all points on the boundary of null-restricted space are limits of core
sequences. Non-core sequences can only converge to the origin, a single
point compared to the continuum of the full boundary. We may now state :

\begin{theorem}[Local Power]
\label{local power theorem}\textit{Assume [A1], [A2], [A3], [A4], [A6] and
[D1], [D3], [D4] hold with the elements }$\mu _{j}^{{}}$\textit{\ of }$\mu $%
\textit{\ taking the T-dependent forms as specified by (\ref{local}). Define 
}%
\begin{eqnarray*}
\tau &\equiv &\dsum\limits_{j=1}^{p}1\{\gamma _{j}=0\}\theta _{j}c_{j} \\
\kappa &\equiv &\dsum\limits_{i=1}^{p}\dsum\limits_{j=1}^{p}1\{\gamma
_{i}=0\}1\{\gamma _{j}=0\}\theta _{i}\theta _{j}v_{ij}
\end{eqnarray*}%
\textit{where }$v_{ij}$\textit{\ denotes the }$(i,j)$-th \textit{element of
variance matrix }$V$\textit{. Assume }$\kappa >0$\textit{.} \textit{Then, as}%
$\ T\ \longrightarrow \infty ,$\textit{\ }%
\begin{equation}
P(Q<\alpha )\longrightarrow \Phi (z_{\alpha }-\kappa ^{-1/2}\tau )\mathit{,}
\label{nor_thm3}
\end{equation}%
\textit{where }$z_{\alpha }$\textit{\ is the }$\alpha $\textit{\ quantile of
the standard normal distribution.}
\end{theorem}

Theorem \ref{local power theorem} implies that the test has power exceeding
size against all core sequences because the composite drift parameter $\tau $
is necessarily negative for such local scenarios. By contrast, tests based
on LFC critical values can be biased against core local sequences tending to
boundary points off the origin. This is easily seen for statistics such as
EV and QLR which are continuous in their arguments. In such cases, local
power under any core sequence (\ref{local}) tends to rejection probability
at the boundary point $\mu =(\gamma _{1},\gamma _{2},...,\gamma
_{p})^{\prime }$. Unless this point is the LFC itself, rejection probability
there will be smaller than that at any LFC point by definition. Hence the
LFC critical value based test is biased against core local alternatives. A
similar argument is given in Hansen (2003, 2005).\medskip

Against non-core local sequences, our test can be biased because a trade-off
comes into force between negative and positive $c_{j}$ as Theorem \ref{local
power theorem} shows. Some degree of local bias is common in multivariate
one-sided tests and exists even in GMS procedures using estimated rather
than LFC test critical values, as noted by Andrews and Soares (2010, p.146,
comment (vi)). However, the exact local direction at which a test exhibits
strength or weakness may vary across tests. Therefore, different tests are
complementary rather than competing. To obtain a formal result, we consider
a local sequence converging to the origin, namely $\gamma _{j}=0$ for $j\in
\{1,2,...,p\}$. Let $c$ denote the vector $(c_{1},c_{2},...,c_{p})^{\prime }$%
. Under such a local scenario, the GMS procedure will asymptotically treat
all inequalities as binding in the critical value calculation. Thus the
asymptotic distribution of the statistic $S(\sqrt{T}\widehat{\mu },\widehat{V%
})$ of Subsection \ref{RL_1} is the same as that of $S(Z+c,V)$ and the test
rejection probability tends to 
\begin{equation}
P(S(Z+c,V)>q_{\alpha })  \label{local bias}
\end{equation}%
where $q_{\alpha }$ is the $(1-\alpha )$ quantile of $S(Z,V)$ under $Z\sim
N(0,V)$. We now present a theorem showing that the test of this paper is
locally most powerful for a non-empty subclass of directions. Let $\theta $\
denote the vector of diagonal elements of the matrix $\Delta $.

\begin{theorem}
\label{local power}\textit{Suppose the variance matrix }$V$\textit{\ is
positive definite and }$\gamma _{j}=0$\textit{\ for }$j\in \{1,2,...,p\}$%
\textit{\ in the local sequence (\ref{local}). Then for every testing
function }$S(.,.)$\textit{\ such that }$P(S(Z,V)>q_{\alpha })=\alpha $%
\textit{\ under }$Z\sim N(0,V)$\textit{, the asymptotic local power in (\ref%
{nor_thm3}) is at least }$\alpha $\textit{\ and is not smaller than (\ref%
{local bias}) when }$c=-\delta V\theta $\textit{\ for any positive scalar }$%
\delta $\textit{.}
\end{theorem}

Depending on the off-diagonal elements of $V$, the local directions $-\delta
V\theta $ can be for either core or non-core sequences.\footnote{%
Note that the vector $-\delta V\theta $ necessarily contains at least one
negative element since $V$ is positive definite, $\theta $ is a positive
vector and $\delta $ is a postive scalar.} Theorem \ref{local power} implies
that along such local alternatives, the present test is not biased and its
limiting local power is not dominated by those of existing tests based on
GMS critical values. Note that the result of Theorem \ref{local power} does
not require specification of particular functional forms of $S(.,.)$. It is
achieved by indirectly exploiting the Neyman-Pearson lemma. Some special
forms are used in Section \ref{nor_s6} for numerical illustration.

\section{Monte Carlo Simulation Studies\label{nor_s6}}

In this section we conduct a series of Monte Carlo simulations to study the
finite sample performance of the test. All tables of simulation results are
placed together at the end of the section.

\subsection{The Specification of Smoothed Indicator\label{nor_s5}}

Our objective is to investigate how well the asymptotic theory of the test
works in finite sample simulations. For this purpose, we choose $\Psi $
functions which are simple, recognized and not contrived. It would be
premature at this stage to undertake a more elaborate exercise to find an
optimal combination of $\Psi (x)$ and $K(T)$. \medskip

For the specification of $\Psi $, the following functions are heuristic
choices that are widely adopted in research on smoothed threshold crossing
models.%
\begin{eqnarray*}
\text{\textit{Normal}} &:&\Psi _{Nor}(x)\equiv 1-\Phi (x) \\
\text{\textit{Logistic}} &:&\Psi _{Log}(x)\equiv (1+\exp (x))^{-1}
\end{eqnarray*}%
Besides $\Psi _{Nor}$ and $\Psi _{Log}$, the following simple choice of $%
\Psi $, mentioned in Section \ref{smoothing}, is also valid.

\begin{equation*}
\text{\textit{Step-at-unity}}:\Psi _{Step}(x)\equiv 1\{x\leq 1\}
\end{equation*}%
As regards the choice of $K(T)$, the following two specifications closely
match tuning parameters used in recent literature on inference of moment
inequality models (See e.g. Chernozhukov et al.\thinspace (2007) and Andrews
and Soares (2010)). These choices are%
\begin{eqnarray*}
\text{\textit{SIC}} &:&K_{SIC}(T)\equiv \sqrt{T/\log (T)} \\
\text{\textit{LIL }} &\mathit{:}&K_{LIL}(T)\equiv \sqrt{T/(2\log \log (T))}
\end{eqnarray*}%
The first name reflects a connection with the Schwarz Information Criterion
(SIC) for model selection and the second with the Law of the Iterated
Logarithm (LIL). \medskip

\subsection{The Simulation Setup}

The simulation experiments are designed as follows. We choose a nominal test
size of $\alpha =0.05$. We use $R=10000$ replications for simulated
rejection probabilities. In each replication, we generate i.i.d.\thinspace
observations $\{x_{t}\}_{t=1}^{T}$ with $T=250$ according to the following
scheme :%
\begin{equation}
x_{t}=\mu +V^{1/2}w_{t}  \label{xt}
\end{equation}%
where $w_{t}$ is a $p$ dimensional random vector whose elements are
i.i.d.\thinspace from distribution $G_{w}.$\medskip

We compute $\widehat{\mu }$ and $\widehat{V}$ as the sample average and
sample variance of the generated data. We take the scalars $\theta _{j}=1/%
\sqrt{v_{jj}}$ and $\widehat{\theta }_{j}=1/\sqrt{\widehat{v}_{jj}}$ where $%
v_{jj}$ and $\widehat{v}_{jj}$ are the $j$th diagonal elements of $V$ and $%
\widehat{V}$ respectively. This simple simulation setup is also adopted by
Andrews and Soares (2010) and Andrews and Barwick (2012) in simulation study
of the GMS tests. For $G_{w}$, we consider three distributions: standard
normal, logistic and $U(-1,2)$, the uniform distribution on the interval $%
[-1,2]$. All of these distributions are centered and scaled such that $%
E(w_{t,j})=0$ and $Var(w_{t,j})=1$ for $j\in \{1,2,...,p\}$. Standard
normality of $G_{w}$ is the benchmark. The logistic distribution has thicker
tails than the normal whilst the support of a uniform distributed random
variate is bounded. The latter two distributions are included to assess the
test performance under finite sample non-normality of $\widehat{\mu }$. For
comparison, we also conduct simulations using the following test statistics:

\begin{eqnarray*}
S_{1} &=&-\min \{\sqrt{T}\widehat{\theta }_{1}\widehat{\mu }_{1},\sqrt{T}%
\widehat{\theta }_{2}\widehat{\mu }_{2},...,\sqrt{T}\widehat{\theta }_{p}%
\widehat{\mu }_{p},0\}, \\
S_{2} &=&\min_{\mu :\mu \geq 0}T(\widehat{\mu }-\mu )^{\prime }\widehat{V}%
^{-1}(\widehat{\mu }-\mu ), \\
S_{3} &=&\dsum\limits_{j=1}^{p}(\min \{\sqrt{T}\widehat{\theta }_{j}\widehat{%
\mu }_{j},0\})^{2}, \\
S_{4} &=&\dsum\limits_{j=1}^{p}[-\sqrt{T}\min (\widehat{\theta }_{j}\widehat{%
\mu }_{j},0)].
\end{eqnarray*}%
\newline

The extreme value form $S_{1}$ is essentially Hansen (2005)'s test statistic
appropriated for testing multiple non-negativity hypotheses. $S_{2}$ is the
classic QLR test statistic. $S_{3}$ is the modified-method-of-moments (MMM)
statistic considered in the literature of moment inequality models (see,
e.g.\thinspace Chernozhukov et al.\thinspace (2007), Romano and Shaikh
(2008), Andrews and Guggenberger (2009) and Andrews and Soares (2010)). $%
S_{4}$ is the raw sum-of-negative-part statistic which can be transformed by
smoothing into the key component of the test of the present paper.\medskip

The critical values for tests based on $S_{1}$ to $S_{4}$ are estimated
using bootstrap coupled with the GMS procedure of the elementwise t-test
type as suggested by Andrews and Soares (2010) and Andrews and Barwick
(2012). We use 10000 bootstrap repetitions for calculation of the GMS test
critical values. The tuning parameter in the GMS procedure is set to be the
SIC or LIL type (Andrews and Soares (2010, p. 131)). For ease of reference,
let $S_{j}(SIC)$ and $S_{j}(LIL)$ denote the GMS test using statistic $S_{j}$
with tuning $SIC$ and $LIL$ respectively. Furthermore, let $Q(\Psi ,K)$
denote the present test implemented with its smoothed indicator specified by 
$\Psi $ and $K$.\medskip

We consider simulation scenarios based on $p\in \{4,6,10\}$. For
multivariate simulation design, we have to be more selective on the
specifications of $\mu $ and $V$ parameters of (\ref{xt}). Concerning the $%
\mu $ vector, we follow a design similar to that previously employed by
Hansen (2005, p. 373) in simulation study of the test size performance. To
be specific, $\mu $ is the $p$ dimensional vector given by%
\begin{equation*}
\mu _{1}^{{}}=0,\mu _{j}^{{}}=\lambda (j-1)/(p-1)\text{ }\ \text{for }p\geq
j\geq 2
\end{equation*}%
where $\lambda \in \{0,0.25,0.5\}$. Note that the $\lambda $ values are
introduced to control the extent to which inequalities satisfying the null
hypothesis are in fact non-binding. Regarding the variance matrix $V$, we
set $V$ to be a Toeplitz matrix with elements $V_{i,j}=\rho ^{j-i}$ for $%
j\geq i,$ where $\rho \in \{0,-0.5,0.5\}$. This greatly simplifies the
specification for off-diagonal elements of $V$ but still allows for presence
of various degrees of both positive and negative correlations.\medskip

For power studies, we consider the $\mu $ vector given by%
\begin{equation}
\mu =-\delta V\theta +\epsilon \widetilde{\mu }
\label{local power simulation}
\end{equation}%
where $\delta \in \{0.15,0.1,0.05\}$, $V$ is the variance matrix given as
above, $\theta =(\theta _{1},\theta _{2},...,\theta _{p})^{\prime }$, $%
\epsilon \in \{0,0.5,0.8\}$ and $\widetilde{\mu }$ is the vector with $%
\widetilde{\mu }_{j}=\delta $ for $1\leq j\leq p/2$ and $\widetilde{\mu }%
_{j}=-\delta $ for $p/2<j\leq p$. For $\epsilon =0$, the design (\ref{local
power simulation}) mimics the local direction as suggested by Theorem \ref%
{local power} under which the test $Q(\Psi ,K)$ is expected to outperform
other tests. When $\epsilon $ is non-zero, the local direction in favor of
the present test is perturbed with another vector $\widetilde{\mu }$
containing mixture of positive and negative elements. Such $\widetilde{\mu }$
may incur power trade-off in light of Theorem \ref{local power theorem} and
thus the perturbation parameter $\epsilon $ controls the degree of deviation
toward $\widetilde{\mu }$ and enables some sensitivity check of test power
performance.

\subsection{Simulation results}

We report the simulated maximum null rejection probability (MNRP) and
average power (AP) for each test. Given $G_{w}$, the maximization for the
MNRP is over all $H_{0}$ compatible combinations of $\mu $ and $\rho $
values whilst given both $G_{w}$ and $\epsilon $, the averaging for AP is
over all $H_{1}$ compatible $\mu $ and $\rho $ configurations. Table 1 lists
the MNRP values in three block columns side by side for the three
specifications of $G_{w}$. The AP values generated by three $\epsilon $
values are then listed separately for each $G_{w}$ in Tables 2, 3 and 4.
\medskip

In Table 1, the primary interest is how close the MNRP values are to the
nominal 5\% significance level, particularly in cases of over-rejecting. In
that respect, we compare the percentage of values not exceeding 0.05, 0.055,
0.06, 0.065. These percentages are about 18, 51, 87, 96 for the 54 $Q(\Psi
,K)$ values and 9, 52, 79, 94 for the 72 values of the GMS tests. Plainly,
the $Q(\Psi ,K)$ test is no more prone to over-rejection than the GMS tests.
A common feature across all tests is that over-rejection tends to increase
with $p$. However, only 2 out of 54 $Q(\Psi ,K)$ entries and 4 out of 72 GMS
entries exceed 0.065. These excesses amount to less than 5\% of a table of
126 simulated entries. \medskip

We now examine the sensitivity of MNRP to the underlying data generating
distribution $G_{w}$. For all tests, Table 1 exhibits little systematic
difference attributable to the three different specifications of $G_{w}$.
These figures suggest that the MNRP results are not sensitive to finite
sample non-normality. Furthermore, for each test, regardless of $G_{w}$,
Table 1 suggests that use of $SIC$ type tuner in place of the $LIL$ can
yield better control of test size. This finding is consistent with the
simulation studies of Andrews and Soares (2010, pp.\thinspace 149-152)
demonstrating that the $SIC$ tuner tends to give better MNRP properties.
Overall, $Q(\Psi _{Step},K_{SIC})$ and $Q(\Psi _{Log},K_{SIC})$ have better
MNRP results among the class of $Q(\Psi ,K)$ tests and their size
performance is comparable to that of the four $SIC$ tuned GMS tests.\medskip

We now turn to Tables 2, 3, 4 giving AP results of the tests. For the
unperturbed direction ($\epsilon =0$), Theorem \ref{local power} of Section %
\ref{power of the test} indicates that the $Q(\Psi ,K)$ test is locally more
powerful than the GMS tests considered in the simulations. Along such local
direction, irrespective of the underlying $G_{w}$, the simulation results
indicate that the $Q(\Psi ,K)$ tests dominate the GMS tests in AP
performance. The GMS QLR test ($S_{2}$) is not far behind. Hansen's test ($%
S_{1}$), which is arguably the most stable in terms of MNRP performance, has
distinctly lower power. But it is still a good performer. For the perturbed
directions ($\epsilon \in \{0.5,0.8\}$), while the $Q(\Psi ,K)$ tests still
outperform the $S_{1}$ tests, they do not generally dominate other versions
of the GMS tests but the AP differences are not large.\medskip

We comment on the comparative performance of the $Q(\Psi ,K)$ tests with the 
$S_{4}$ tests. Their comparison is of particular interest since the present
test essentially attempts to smooth the statistic $S_{4}$. The smoothed
version is less costly in computation because its critical value is obtained
without resampling. We compare $S_{4}(SIC)$ with $Q(\Psi _{Step},K_{SIC})$
and $Q(\Psi _{Log},K_{SIC})$. The simulation results suggest that the $%
Q(\Psi _{Step},K_{SIC})$ and $Q(\Psi _{Log},K_{SIC})$ tests have similar
degree of size control as $S_{4}(SIC)$. Against the alternative hypothesis, $%
Q(\Psi _{Log},K_{SIC})$ has slightly larger power than $S_{4}(SIC)$ in all
27 cases while $Q(\Psi _{Step},K_{SIC})$ outperforms $S_{4}(SIC)$ in 18 out
of the 27 cases. These findings suggest that implementational advantage of
the present test based on smoothing does not appear to be achieved at the
cost of test performance. \medskip

Perusing all the other entries in Tables 2, 3, 4, it seems that the
different variants of the $Q(\Psi ,K)$ test perform quite similarly to one
another retaining power well in excess of 0.73 throughout. What these
results illustrate is that the $Q(\Psi ,K)$ test has identifiable directions
of strength as indicated theoretically by this paper. Given the simulation
results above, the $Q(\Psi _{Step},K_{SIC})$ and $Q(\Psi _{Log},K_{SIC})$
tests work at least as well as other $Q(\Psi ,K)$ versions examined here but
have better size performance. Hence while $K_{SIC}$ is the preferred tuner,
both $\Psi _{Step}$ and $\Psi _{Log}$ are the recommended smoothers.

\bigskip

\begin{center}
\begin{tabular}{l|ccc|ccc|ccc}
\multicolumn{10}{c}{Table 1 : Simulated Maximum Null Rejection Probability
for $T=250$} \\ \hline\hline
DGP $G_{w}$ & \multicolumn{3}{|c|}{$N(0,1)$} & \multicolumn{3}{|c|}{$%
Logistic $} & \multicolumn{3}{|c}{$U(-1,2)$} \\ \hline
Number of inequalities & 4 & 6 & 10 & 4 & 6 & 10 & 4 & 6 & 10 \\ \hline
$Q(\Psi _{Step},K_{SIC})$ & .049 & .056 & .055 & .052 & .054 & .056 & .051 & 
.052 & .055 \\ 
$Q(\Psi _{Log},K_{SIC})$ & .046 & .053 & .055 & .046 & .054 & .057 & .048 & 
.052 & .058 \\ 
$Q(\Psi _{Nor},K_{SIC})$ & .050 & .059 & .061 & .050 & .058 & .063 & .050 & 
.056 & .063 \\ 
$Q(\Psi _{Step},K_{LIL})$ & .051 & .059 & .059 & .053 & .056 & .059 & .051 & 
.053 & .057 \\ 
$Q(\Psi _{Log},K_{LIL})$ & .049 & .056 & .057 & .048 & .057 & .060 & .048 & 
.053 & .059 \\ 
$Q(\Psi _{Nor},K_{LIL})$ & .054 & .062 & .065 & .052 & .059 & .066 & .053 & 
.058 & .066 \\ 
$S_{1}(SIC)$ & .050 & .052 & .054 & .049 & .052 & .053 & .051 & .052 & .053
\\ 
$S_{2}(SIC)$ & .050 & .054 & .053 & .052 & .055 & .054 & .050 & .050 & .054
\\ 
$S_{3}(SIC)$ & .050 & .056 & .052 & .050 & .051 & .057 & .052 & .052 & .056
\\ 
$S_{4}(SIC)$ & .051 & .058 & .054 & .053 & .054 & .057 & .052 & .055 & .058
\\ 
$S_{1}(LIL)$ & .053 & .055 & .055 & .051 & .054 & .056 & .054 & .054 & .056
\\ 
$S_{2}(LIL)$ & .058 & .061 & .061 & .059 & .063 & .063 & .058 & .058 & .061
\\ 
$S_{3}(LIL)$ & .056 & .061 & .057 & .055 & .058 & .065 & .058 & .058 & .064
\\ 
$S_{4}(LIL)$ & .059 & .068 & .066 & .060 & .064 & .070 & .061 & .065 & .070
\\ \hline
\end{tabular}

\bigskip

\bigskip

\begin{tabular}{l|ccc|ccc|ccc}
\multicolumn{10}{c}{Table 2 : Simulated Average Power for $T=250$, $%
G_{w}=N(0,1)$} \\ \hline\hline
& \multicolumn{3}{|c|}{$\epsilon =0$} & \multicolumn{3}{|c|}{$\epsilon =0.5$}
& \multicolumn{3}{|c}{$\epsilon =0.8$} \\ \hline
Number of inequalities & 4 & 6 & 10 & 4 & 6 & 10 & 4 & 6 & 10 \\ \hline
\multicolumn{1}{l|}{$Q(\Psi _{Step},K_{SIC})$} & .770 & .837 & .900 & .773 & 
.840 & .904 & .783 & .849 & .909 \\ 
\multicolumn{1}{l|}{$Q(\Psi _{Log},K_{SIC})$} & .754 & .827 & .893 & .783 & 
.849 & .910 & .813 & .872 & .927 \\ 
\multicolumn{1}{l|}{$Q(\Psi _{Nor},K_{SIC})$} & .741 & .814 & .882 & .780 & 
.845 & .906 & .817 & .875 & .928 \\ 
\multicolumn{1}{l|}{$Q(\Psi _{Step},K_{LIL})$} & .752 & .822 & .886 & .761 & 
.830 & .895 & .780 & .847 & .906 \\ 
\multicolumn{1}{l|}{$Q(\Psi _{Log},K_{LIL})$} & .748 & .821 & .888 & .781 & 
.847 & .908 & .815 & .874 & .928 \\ 
\multicolumn{1}{l|}{$Q(\Psi _{Nor},K_{LIL})$} & .734 & .807 & .875 & .778 & 
.844 & .903 & .819 & .876 & .928 \\ 
$S_{1}(SIC)$ & .593 & .626 & .650 & .699 & .728 & .761 & .774 & .803 & .831
\\ 
$S_{2}(SIC)$ & .714 & .781 & .847 & .784 & .844 & .901 & .834 & .887 & .937
\\ 
$S_{3}(SIC)$ & .678 & .735 & .793 & .750 & .804 & .858 & .805 & .854 & .899
\\ 
$S_{4}(SIC)$ & .730 & .794 & .855 & .767 & .830 & .886 & .808 & .864 & .913
\\ 
$S_{1}(LIL)$ & .594 & .626 & .650 & .700 & .729 & .762 & .776 & .805 & .832
\\ 
$S_{2}(LIL)$ & .716 & .782 & .848 & .785 & .846 & .903 & .836 & .889 & .939
\\ 
$S_{3}(LIL)$ & .678 & .736 & .794 & .751 & .805 & .860 & .808 & .856 & .902
\\ 
$S_{4}(LIL)$ & .732 & .795 & .857 & .769 & .833 & .889 & .811 & .868 & .916
\\ \hline
\end{tabular}

\bigskip

\bigskip

\begin{tabular}{l|ccc|ccc|ccc}
\multicolumn{10}{c}{Table 3 : Simulated Average Power for $T=250$, $%
G_{w}=Logistic$} \\ \hline\hline
& \multicolumn{3}{|c|}{$\epsilon =0$} & \multicolumn{3}{|c|}{$\epsilon =0.5$}
& \multicolumn{3}{|c}{$\epsilon =0.8$} \\ \hline
Number of inequalities & 4 & 6 & 10 & 4 & 6 & 10 & 4 & 6 & 10 \\ \hline
\multicolumn{1}{l|}{$Q(\Psi _{Step},K_{SIC})$} & .772 & .839 & .900 & .774 & 
.841 & .903 & .781 & .850 & .910 \\ 
\multicolumn{1}{l|}{$Q(\Psi _{Log},K_{SIC})$} & .757 & .828 & .893 & .785 & 
.851 & .910 & .813 & .875 & .929 \\ 
\multicolumn{1}{l|}{$Q(\Psi _{Nor},K_{SIC})$} & .744 & .815 & .882 & .781 & 
.847 & .906 & .817 & .878 & .930 \\ 
\multicolumn{1}{l|}{$Q(\Psi _{Step},K_{LIL})$} & .753 & .824 & .886 & .763 & 
.831 & .894 & .779 & .848 & .908 \\ 
\multicolumn{1}{l|}{$Q(\Psi _{Log},K_{LIL})$} & .751 & .823 & .888 & .783 & 
.849 & .908 & .815 & .876 & .930 \\ 
\multicolumn{1}{l|}{$Q(\Psi _{Nor},K_{LIL})$} & .738 & .808 & .874 & .780 & 
.845 & .904 & .819 & .878 & .930 \\ 
$S_{1}(SIC)$ & .599 & .629 & .651 & .697 & .729 & .762 & .775 & .803 & .831
\\ 
$S_{2}(SIC)$ & .718 & .782 & .847 & .784 & .845 & .901 & .834 & .889 & .938
\\ 
$S_{3}(SIC)$ & .681 & .737 & .794 & .750 & .803 & .858 & .806 & .855 & .901
\\ 
$S_{4}(SIC)$ & .734 & .795 & .854 & .768 & .830 & .886 & .807 & .866 & .915
\\ 
$S_{1}(LIL)$ & .600 & .629 & .651 & .699 & .730 & .763 & .777 & .805 & .833
\\ 
$S_{2}(LIL)$ & .719 & .784 & .849 & .786 & .846 & .903 & .837 & .891 & .940
\\ 
$S_{3}(LIL)$ & .682 & .738 & .796 & .751 & .805 & .861 & .808 & .857 & .903
\\ 
$S_{4}(LIL)$ & .735 & .797 & .856 & .771 & .833 & .889 & .811 & .869 & .919
\\ \hline
\end{tabular}

\bigskip

\bigskip

\begin{tabular}{l|ccc|ccc|ccc}
\multicolumn{10}{c}{Table 4 : Simulated Average Power for $T=250$, $%
G_{w}=U(-1,2)$} \\ \hline\hline
& \multicolumn{3}{|c|}{$\epsilon =0$} & \multicolumn{3}{|c|}{$\epsilon =0.5$}
& \multicolumn{3}{|c}{$\epsilon =0.8$} \\ \hline
Number of inequalities & 4 & 6 & 10 & 4 & 6 & 10 & 4 & 6 & 10 \\ \hline
\multicolumn{1}{l|}{$Q(\Psi _{Step},K_{SIC})$} & .769 & .837 & .899 & .775 & 
.842 & .902 & .782 & .849 & .908 \\ 
\multicolumn{1}{l|}{$Q(\Psi _{Log},K_{SIC})$} & .754 & .826 & .892 & .785 & 
.850 & .910 & .812 & .874 & .926 \\ 
\multicolumn{1}{l|}{$Q(\Psi _{Nor},K_{SIC})$} & .741 & .813 & .880 & .781 & 
.846 & .906 & .817 & .876 & .927 \\ 
\multicolumn{1}{l|}{$Q(\Psi _{Step},K_{LIL})$} & .752 & .821 & .885 & .763 & 
.832 & .894 & .779 & .847 & .907 \\ 
\multicolumn{1}{l|}{$Q(\Psi _{Log},K_{LIL})$} & .749 & .820 & .886 & .784 & 
.848 & .908 & .815 & .876 & .927 \\ 
\multicolumn{1}{l|}{$Q(\Psi _{Nor},K_{LIL})$} & .735 & .806 & .873 & .780 & 
.844 & .903 & .819 & .878 & .928 \\ 
$S_{1}(SIC)$ & .594 & .623 & .652 & .698 & .727 & .758 & .773 & .801 & .830
\\ 
$S_{2}(SIC)$ & .715 & .778 & .846 & .784 & .843 & .900 & .834 & .887 & .937
\\ 
$S_{3}(SIC)$ & .678 & .733 & .793 & .749 & .803 & .858 & .805 & .854 & .899
\\ 
$S_{4}(SIC)$ & .730 & .793 & .852 & .768 & .831 & .886 & .807 & .866 & .914
\\ 
$S_{1}(LIL)$ & .594 & .623 & .652 & .699 & .728 & .759 & .775 & .803 & .831
\\ 
$S_{2}(LIL)$ & .716 & .780 & .848 & .785 & .845 & .902 & .836 & .889 & .939
\\ 
$S_{3}(LIL)$ & .679 & .734 & .794 & .751 & .805 & .860 & .807 & .857 & .901
\\ 
$S_{4}(LIL)$ & .731 & .794 & .853 & .770 & .833 & .889 & .811 & .869 & .918
\\ \hline
\end{tabular}
\end{center}

\section{Conclusions\label{nor_s8}}

This paper develops a test of multiple inequality hypotheses whose
implementation does not require computationally intensive procedures. The
test is based on origin-smooth approximation of indicators underlying the
sum-of-negative-part statistic. This yields a simply structured statistic
whose asymptotic distribution, whenever non-degenerate, is normal under the
null hypothesis. Hence test critical values can be fixed ex ante and are
essentially based on the unit normal distribution. Moreover, the test is
applicable under weak assumptions allowing for estimator covariance
singularity\medskip .

We have proved that the size of the test is asymptotically exact in the
uniform sense. The test is consistent against all fixed alternative
hypotheses. We have derived a local power function and used it to
demonstrate that the test is unbiased against a wide class of local
alternatives. We have also provided a new theoretical result pinpointing
directions of alternatives for which the test is locally most
powerful.\medskip

We have performed simulations which illustrate the potential of the test to
be of practical inferential value along with simplicity and speed. These
simulations, carried out for a range of $p$ values, also shed light on the
choice of smoothed indicator. They suggest that when coupled with the SIC
type tuner, both the logistic and the step-at-unity smoothers perform well
in finite samples. These are the recommended choices for test
implementation. The simulation study also compares the test of this paper
with several different tests which estimate critical values using the GMS
procedure. We find that the test appears to be a viable complement to the
GMS critical value estimation methodology.

\newpage

\appendix%

\section{Supplementary Derivation of $\Lambda _{T}(\widehat{\protect\mu }%
_{j},\widehat{v}_{jj})$}

The term $\Lambda _{T}(\widehat{\mu }_{j},\widehat{v}_{jj})$ acts as an
approximation for the expectation of $[\Psi _{T}(\widehat{\mu }_{j})-\Psi
(0)]\sqrt{T}\widehat{\mu }_{j}$ evaluated at $\mu _{j}=0$. Under regularity
condition [D1], when $\mu _{j}=0$, the distribution of $\sqrt{T}\widehat{\mu 
}_{j}$ for $T$ sufficiently large is approximately normal with mean zero and
variance $v_{jj}$. Let $X$ denote any scalar random variable distributed as $%
N(0,c)$. Define $h_{T}\equiv K(T)/\sqrt{T}$. Given (\ref{psiT}), $\Lambda
_{T}(\widehat{\mu }_{j},\widehat{v}_{jj})$ is thus constructed to
approximate $E((\Psi (h_{T}X)-\Psi (0))X)=E(\Psi (h_{T}X)X)$ with $c=v_{jj}$%
. In what follows, we take as read the notation and definitions stated
between equations (\ref{psiT}) and (\ref{gammaT}).\medskip

Define $a_{0}\equiv -\infty $ and $a_{n+1}\equiv \infty $. Let $\phi $
denote the standard normal density function. Note that 
\begin{eqnarray}
&&E(\Psi (h_{T}X)X)  \notag \\
&=&\sum_{i=1}^{n+1}\int_{a_{i-1}/h_{T}}^{a_{i}/h_{T}}\Psi (h_{T}x)x\phi (x/%
\sqrt{c})/\sqrt{c}dx  \notag \\
&=&\sqrt{c}\left[ \sum_{i=1}^{n+1}\int_{a_{i-1}/h_{T}}^{a_{i}/h_{T}}h_{T}%
\psi (h_{T}x)\phi (x/\sqrt{c})dx-\sum_{i=1}^{n}(\Psi (a_{i}^{-})-\Psi
(a_{i}^{+}))\phi (\frac{a_{i}}{h_{T}\sqrt{c}})\right]  \label{c_1} \\
&=&ch_{T}E(\widetilde{\psi }(h_{T}X))-\sqrt{c}\sum_{i=1}^{n}(\Psi
(a_{i}^{-})-\Psi (a_{i}^{+}))\phi (\frac{a_{i}}{h_{T}\sqrt{c}})  \label{c_4}
\end{eqnarray}%
where (\ref{c_1}) follows from integration by parts and re-arrangement of
terms in the sum and (\ref{c_4}) follows by using [A2] which implies $%
\widetilde{\psi }(x)=\psi (x)$ almost everywhere. Taking $c=v_{jj}$ and
plugging in the parameter estimates, we hence construct $\Lambda _{T}(%
\widehat{\mu }_{j},\widehat{v}_{jj})$ as%
\begin{equation}
\Lambda _{T}(\widehat{\mu }_{j},\widehat{v}_{jj})\equiv \widehat{v}_{jj}%
\widetilde{\psi }(K(T)\widehat{\mu }_{j})K(T)/\sqrt{T}-\sqrt{\widehat{v}_{jj}%
}\sum_{i=1}^{n}(\Psi (a_{i}^{-})-\Psi (a_{i}^{+}))\phi (\frac{a_{i}\sqrt{T}}{%
\sqrt{\widehat{v}_{jj}}K(T)}).  \label{gammaT term}
\end{equation}%
We now comment on the derivative term in the expression (\ref{gammaT term}).
Since $h_{T}$ goes to zero as $T$ increases, $E(\widetilde{\psi }(h_{T}X))$
tends to $\psi (0)$ by Assumption [A2] and the Dominated Convergence
Theorem. The limit value $\psi (0)$ also coincides with the probability
limit of $\widetilde{\psi }(K(T)\widehat{\mu }_{j})$ for the case $\mu
_{j}=0 $. Hence, we use $\widetilde{\psi }(K(T)\widehat{\mu }_{j})$ instead
of $E(\widetilde{\psi }(h_{T}X))$ to account for the slope effect,\footnote{%
By taking $X\sim N(0,c)$ with $c=\widehat{v}_{jj}$, $E(\widetilde{\psi }%
(h_{T}X))$ can be computed using numerical integral as 
\begin{equation*}
\int_{-\infty }^{\infty }\widetilde{\psi }(h_{T}x)\phi (x/\sqrt{\widehat{v}%
_{jj}})/\sqrt{\widehat{v}_{jj}}dx.
\end{equation*}%
} thus allowing the derivative term to depend on the estimate $\widehat{\mu }%
_{j}$. This has the advantage that for non-zero valued $\mu _{j}$, $%
\widetilde{\psi }(K(T)\widehat{\mu }_{j})$ itself also tends to zero and
hence yields faster convergence of $\Lambda _{T}$ to zero when the function $%
\Psi $ further has the properties of $\lim_{x\longrightarrow -\infty }\psi
(x)=\lim_{x\longrightarrow \infty }\psi (x)=0$. Specifications of $\Psi $
satisfying these properties are numerous, including the logistic and the
normal smoothers given in Section \ref{nor_s5}.

\section{Proofs of Theoretical Results}

The section presents proofs of all theoretical results stated in the paper.
Proofs of Theorems 1, 3, 4 and 5 (pointwise asymptotics and local power)
along with preliminary Lemmas 1, 2 and 3 are presented in Subsections \ref%
{B1} - \ref{B7}. Proofs of Lemma 4 providing a sufficient condition for
Assumption [U2] and Theorem 2 (uniform asymptotics) are given separately in
Subsections \ref{B8} and \ref{B9} of the Appendix. \textit{\medskip }

Recall that $J$ denotes the set $\{1,2,...,p\}$ and the sets $A$, $M$, and $%
B $ are defined as%
\begin{equation*}
A\equiv \{j\in J:\mu _{j}>0\},\text{ }M\equiv \{j\in J:\mu _{j}=0\},\text{ }%
B\equiv \{j\in J:\mu _{j}<0\}.
\end{equation*}

\subsection{Probability Limits of the Smoothed Indicator\label{B1}}

We first prove a lemma that states the probability limits of the smoothed
indicator $\Psi _{T}(\widehat{\theta }_{j}\widehat{\mu }_{j})$, which will
be referred to in the proofs of some theorems in this paper.\bigskip

\begin{lemma}[Probability Limits of the Smoothed Indicator ]
\label{Ch_nor_lemma}
\end{lemma}

\textit{Assume [D1] and [D4]. Then the following results are valid as }$%
T\longrightarrow \infty .$\textit{\medskip }

(1)\textit{\ \ \ If }$j\in A$ \textit{and [A1], [A3], [A6] hold, then }$%
\sqrt{T}\Psi _{T}(\widehat{\theta }_{j}\widehat{\mu }_{j})\overset{p}{%
\longrightarrow }$\textit{\ }$0\smallskip .$

(2)\textit{\ \ \ If }$j\in M$ \textit{and [A2], [A4] hold, then }$\Psi _{T}(%
\widehat{\theta }_{j}\widehat{\mu }_{j})\overset{p}{\longrightarrow }$%
\textit{\ }$\Psi (0).\smallskip $

(3) \textit{\ \ If }$j\in B$ \textit{and [A1], [A3], [A5] hold, then }$\Psi
_{T}(\widehat{\theta }_{j}\widehat{\mu }_{j})\overset{p}{\longrightarrow }$%
\textit{\ }$1.$

\bigskip

\begin{proof}
To show part (1), for $\varepsilon >0$ and for $\eta >0$, we want to find
some $\overline{T}(\varepsilon ,\eta )>0$ such that for $T>$ $\overline{T}%
(\varepsilon ,\eta )$, 
\begin{equation*}
P(\sqrt{T}\Psi _{T}(\widehat{\theta }_{j}\widehat{\mu }_{j})\leq \varepsilon
)\geq 1-\eta \text{.}
\end{equation*}%
By [D1] and [D4], we have $\widehat{\theta }_{j}\widehat{\mu }_{j}\overset{p}%
{\longrightarrow }$\textit{\ }$\theta _{j}\mu _{j}$, which is strictly
positive for $j\in A$. Then there is a $T_{1}(\eta )$ such that for $%
T>T_{1}(\eta ),$%
\begin{equation*}
P(\theta _{j}\mu _{j}/2\leq \widehat{\theta }_{j}\widehat{\mu }_{j}\leq
3\theta _{j}\mu _{j}/2)\geq 1-\eta .
\end{equation*}%
Therefore, by [A1] and [A3] we have 
\begin{eqnarray*}
1-\eta &\leq &P(\Psi _{T}(3\theta _{j}\mu _{j}/2)\leq \Psi _{T}(\widehat{%
\theta }_{j}\widehat{\mu }_{j})\leq \Psi _{T}(\theta _{j}\mu _{j}/2)) \\
&\leq &P(\Psi _{T}(\widehat{\theta }_{j}\widehat{\mu }_{j})\leq \Psi
_{T}(\theta _{j}\mu _{j}/2)) \\
&\leq &P(\sqrt{T}\Psi _{T}(\widehat{\theta }_{j}\widehat{\mu }_{j})\leq 
\sqrt{T}\Psi _{T}(\theta _{j}\mu _{j}/2))
\end{eqnarray*}%
where the first inequality follows because $\Psi $ is a non-increasing
function. [A6] implies that $\sqrt{T}\Psi _{T}(\theta _{j}\mu
_{j}/2)\longrightarrow 0$ as $T\longrightarrow \infty $. Therefore, there is
some $T_{2}(\varepsilon )$ such that for $T>T_{2}(\varepsilon ),$ $\sqrt{T}%
\Psi _{T}(\theta _{j}\mu _{j}/2)<\varepsilon $. Combining all these results,
part (1) in this lemma follows by choosing $\overline{T}(\varepsilon ,\eta
)=\max (T_{1}(\eta ),T_{2}(\varepsilon )).\medskip $

To show part (2), note that If $j\in M$, by [D1] and [D4], we have $\sqrt{T}%
\widehat{\theta }_{j}\widehat{\mu }_{j}=Op(1)$. By [A4], $K(T)/\sqrt{T}=o(1)$
so that $K(T)\widehat{\theta }_{j}\widehat{\mu }_{j}\overset{p}{%
\longrightarrow }$\textit{\ }$0$. By [A2], $\Psi $ is continuous at origin.
Therefore, part (2) follows from the application of the continuous mapping
theorem.\medskip

To show part (3), for $\varepsilon >0$ and for $\eta >0$, we want to find
some $\overline{T}(\varepsilon ,\eta )>0$ such that for $T>$ $\overline{T}%
(\varepsilon ,\eta )$, 
\begin{equation*}
P(1-\varepsilon \leq \Psi _{T}(\widehat{\theta }_{j}\widehat{\mu }_{j})\leq
1+\varepsilon )\geq 1-\eta \text{.}
\end{equation*}%
Following the proof given in part (1), we have that there is a $T_{1}(\eta )$
such that for $T>T_{1}(\eta )$ 
\begin{eqnarray*}
1-\eta &\leq &P(\theta _{j}\mu _{j}/2\leq \widehat{\theta }_{j}\widehat{\mu }%
_{j}\leq 3\theta _{j}\mu _{j}/2) \\
&\leq &P(\Psi _{T}(3\theta _{j}\mu _{j}/2)\leq \Psi _{T}(\widehat{\theta }%
_{j}\widehat{\mu }_{j})\leq \Psi _{T}(\theta _{j}\mu _{j}/2))\text{.}
\end{eqnarray*}%
Note that if \textit{\ }$j\in B$, then $\theta _{j}\mu _{j}<0$ and thus by
[A5], $\Psi _{T}(\theta _{j}\mu _{j}/2)\longrightarrow 1$ and $\Psi
_{T}(3\theta _{j}\mu _{j}/2)\longrightarrow 1$. Then there is some $%
T_{3}(\varepsilon )$ such that for $T>T_{3}(\varepsilon )$, $\Psi
_{T}(\theta _{j}\mu _{j}/2)\leq 1+\varepsilon $ and $\Psi _{T}(3\theta
_{j}\mu _{j}/2)\geq 1-\varepsilon $. Therefore, part (3) follows by choosing 
$\overline{T}(\varepsilon ,\eta )=\max (T_{1}(\eta ),T_{3}(\varepsilon )).$
\end{proof}

\subsection{Asymptotic Properties of $\protect\sqrt{T}\Psi _{T}(\widehat{%
\protect\theta }_{j}\widehat{\protect\mu }_{j})\widehat{\protect\theta }_{j}%
\widehat{\protect\mu }_{j}$}

Based on Lemma \ref{Ch_nor_lemma}, we derive the asymptotic properties of
the components corresponding to \textit{\ }$j\in A,$ \textit{\ }$j\in M,$ 
\textit{\ }$j\in B$ of the sum $\sum_{j\in J}\sqrt{T}\Psi _{T}(\widehat{%
\theta }_{j}\widehat{\mu }_{j})\widehat{\theta }_{j}\widehat{\mu }_{j}$. The
results are stated in the following lemma.\bigskip

\begin{lemma}[Asymptotic Properties of $\protect\sqrt{T}\Psi _{T}(\widehat{%
\protect\theta }_{j}\widehat{\protect\mu }_{j})\widehat{\protect\theta }_{j}%
\widehat{\protect\mu }_{j}$]
\label{Ch_nor_lemma2}
\end{lemma}

\textit{Let }$v_{jj}$\textit{\ denote the }$j$\textit{th diagonal element of 
}$V$\textit{. Assume [D1] and [D4]. Then the following results are valid as }%
$T\longrightarrow \infty .$\textit{\medskip }

(i)\textit{\ \ \ \ \ If }$j\in A$ \textit{and [A1], [A3], [A6] hold, then }$%
\sqrt{T}\Psi _{T}(\widehat{\theta }_{j}\widehat{\mu }_{j})\widehat{\theta }%
_{j}\widehat{\mu }_{j}\overset{p}{\longrightarrow }$\textit{\ }$0\smallskip
. $

(ii)\textit{\ \ \ \ If }$j\in M$ \textit{and [A2], [A4] hold, then }$\sqrt{T}%
\Psi _{T}(\widehat{\theta }_{j}\widehat{\mu }_{j})\widehat{\theta }_{j}%
\widehat{\mu }_{j}\overset{d}{\longrightarrow }$\textit{\ }$N(0,(\Psi
(0)\theta _{j})^{2}v_{jj})\smallskip .$

(iii) \textit{\ \ If }$j\in B$ \textit{and [A1], [A3], [A5] hold, then }$%
\sqrt{T}\Psi _{T}(\widehat{\theta }_{j}\widehat{\mu }_{j})\widehat{\theta }%
_{j}\widehat{\mu }_{j}\overset{p}{\longrightarrow }$\textit{\ }$-\infty .$

\bigskip

\begin{proof}
Note that part (i) follows from [D1], [D4] and part (1) of Lemma \ref%
{Ch_nor_lemma}. To show part (ii), by [D1] and [D4], if $j\in M$\textit{, }%
we have that $\sqrt{T}\widehat{\theta }_{j}\widehat{\mu }_{j}\overset{d}{%
\longrightarrow }N(0,\theta _{j}^{2}v_{jj})$. Therefore, part (ii) follows
by applying part (2) of Lemma \ref{Ch_nor_lemma}. To show part (iii), note
that for $j\in B,$ 
\begin{equation}
\sqrt{T}\Psi _{T}(\widehat{\theta }_{j}\widehat{\mu }_{j})\widehat{\theta }%
_{j}\widehat{\mu }_{j}=\Psi _{T}(\widehat{\theta }_{j}\widehat{\mu }_{j})%
\sqrt{T}\widehat{\theta }_{j}(\widehat{\mu }_{j}-\mu _{j})+\Psi _{T}(%
\widehat{\theta }_{j}\widehat{\mu }_{j})\sqrt{T}\widehat{\theta }_{j}\mu
_{j}.  \label{norm_1}
\end{equation}%
Therefore, part (iii) follows from the fact that by [D1], [D4] and part (3)
of Lemma \ref{Ch_nor_lemma}, the first term on the right hand side of (\ref%
{norm_1}) is $Op(1)$ and the second term goes to $-\infty $ in probability.
\end{proof}

\subsection{Asymptotic Properties of $\Lambda _{T}(\widehat{\protect\theta }%
_{j}\widehat{\protect\mu }_{j},\widehat{\protect\theta }_{j}^{2}\widehat{v}%
_{jj})$}

The following lemma states the asymptotic properties of the adjustment term $%
\Lambda _{T}(\widehat{\theta }_{j}\widehat{\mu }_{j},\widehat{\theta }%
_{j}^{2}\widehat{v}_{jj})$ defined by (\ref{gammaT}).\bigskip

\begin{lemma}[Asymptotic Properties of $\Lambda _{T}(\widehat{\protect\theta 
}_{j}\widehat{\protect\mu }_{j},\widehat{\protect\theta }_{j}^{2}\widehat{v}%
_{jj})$]
\label{Lemma 3}
\end{lemma}

\textit{Assume [A1], [A2], [A4], [D3] and [D4]. Then for }$j\in J$\textit{, }%
$\Lambda _{T}(\widehat{\theta }_{j}\widehat{\mu }_{j},\widehat{\theta }%
_{j}^{2}\widehat{v}_{jj})\overset{p}{\longrightarrow }$\textit{\ }$0$\textit{%
.}

\bigskip

\begin{proof}
By [A1] and [A2] and the properties of standard normal density function, we
find that%
\begin{equation*}
\left\vert \Lambda _{T}(\widehat{\theta }_{j}\widehat{\mu }_{j},\widehat{%
\theta }_{j}^{2}\widehat{v}_{jj})\right\vert \leq \widehat{\theta }_{j}^{2}%
\widehat{v}_{jj}\frac{K(T)}{\sqrt{T}}\left[ b_{\Psi }+\sqrt{2\widehat{\theta 
}_{j}^{2}\widehat{v}_{jj}\pi ^{-1}}\frac{K(T)}{\sqrt{T}}%
\sum_{i=1}^{n}a_{i}^{-2}\right]
\end{equation*}%
where $b_{\Psi }$ denotes the finite positive bound on the derivative of $%
\Psi $ given in Assumption [A2]. Note that [A2] also implies $a_{i}^{2}>0$
for each $i$. By [A4], [D3] and [D4], the right-hand side of the inequality
above is $o_{p}(1)$ and thus Lemma \ref{Lemma 3} follows.
\end{proof}

\subsection{Proof of Theorem 1\label{pf_nor_thm1}}

\textit{Proof of part (1) :\medskip }

By Lemma \ref{Lemma 3} and under $H_{0}$, the quantity $Q_{1}$ may be
written as%
\begin{equation*}
Q_{1}=\dsum\limits_{j\in A}\sqrt{T}\Psi _{T}(\widehat{\theta }_{j}\widehat{%
\mu }_{j})\widehat{\theta }_{j}\widehat{\mu }_{j}+\dsum\limits_{j\in M}\sqrt{%
T}\Psi _{T}(\widehat{\theta }_{j}\widehat{\mu }_{j})\widehat{\theta }_{j}%
\widehat{\mu }_{j}+o_{p}(1)
\end{equation*}%
which, by part (i) of Lemma \ref{Ch_nor_lemma2}, is asymptotically
equivalent in probability to merely%
\begin{equation*}
\dsum\limits_{j\in M}\sqrt{T}\Psi _{T}(\widehat{\theta }_{j}\widehat{\mu }%
_{j})\widehat{\theta }_{j}\widehat{\mu }_{j}.
\end{equation*}%
which, by [D1], [D2], [D4] and part (2) of Lemma \ref{Ch_nor_lemma}, is
asymptotically normal with mean zero and strictly positive variance equal to 
$\Psi (0)^{2}\omega _{M}$ where $\omega _{M}\equiv d_{M}^{\prime }\Delta
V\Delta d_{M}$ in which $d_{M}$ denotes the $p$ dimensional vector whose $j$%
th element is unity for $j\in M$ but zero for $j\notin M$. Using similar
arguments along with [D3], we also find that%
\begin{equation*}
Q_{2}\equiv \sqrt{\widehat{\Psi }^{\prime }\widehat{\Delta }\widehat{V}%
\widehat{\Delta }\widehat{\Psi }}\overset{p}{\longrightarrow }\Psi (0)\omega
_{M}^{1/2}.
\end{equation*}%
From these results about $Q_{1}$ and $Q_{2}$ and the definition (\ref{q}) of 
$Q$, we conclude that $Q$ equals to $\Phi (Q_{1}/Q_{2})$ with probability
tending to 1 as $T\longrightarrow \infty $ and thus $Q\overset{d}{%
\longrightarrow }U(0,1)$.

\bigskip

\textit{Proof of part (2) :\medskip }

When $M$ is empty yet $H_{0}$ holds, only the sums taken for $j\in A$ remain
in the definitions of $Q_{1}$ and $Q_{2}$ hence the following analysis is
confined to $j\in A$. We distinguish between smoothed indicators which are
such that $\Psi _{T}(x)=0$ for all $T$ sufficiently large when $x>0$ and
smoothed indicators such that $\Psi _{T}(x)$ remains strictly positive for $%
x>0$ for all $T$. In the former case, part (1) of Lemma \ref{Ch_nor_lemma}
implies that $P(\Psi _{T}(\widehat{\theta }_{j}\widehat{\mu }%
_{j})=0)\longrightarrow 1$ for $j\in A$ and hence $P(Q_{2}=0)\longrightarrow
1$ and thus $P(Q=1)\longrightarrow 1.$\medskip

Now we consider the latter case where $\Psi _{T}(x)>0$ for $x>0$ regardless
of $T$. This happens for everywhere positive $\Psi $ functions. Then the
quantity $\widehat{\Upsilon }_{j}\equiv \widehat{\theta }_{j}\Psi _{T}(%
\widehat{\theta }_{j}\widehat{\mu }_{j})$ is almost surely strictly positive
for all $j\in A$. By eigenvalue theory, for all $T$,%
\begin{equation}
Q_{2}\leq \sqrt{\widehat{\lambda }_{\max }\dsum\limits_{j\in A}\widehat{%
\Upsilon }_{j}^{2}}\leq \sqrt{p\widehat{\lambda }_{\max }}\max_{j\in A}\{%
\widehat{\Upsilon }_{j}\}  \label{inequality of Q2}
\end{equation}%
where $\widehat{\lambda }_{\max }$ is the largest eigenvalue of $\widehat{V}$%
. Note that (\ref{inequality of Q2}) holds even if $Q_{2}=0$, which under
current scenario could only happen because of singularity of $\widehat{V}$
and $V$. However, when $P(Q_{2}=0)\longrightarrow 1$, we have $%
P(Q=1)\longrightarrow 1$ and thus part (2) of the theorem follows.\medskip

Note that for\textit{\ }$j\in J$, equation (\ref{gammaT}) and Assumptions
[A1] and [A2] imply that the term $\Lambda _{T}(\widehat{\theta }_{j}%
\widehat{\mu }_{j},\widehat{\theta }_{j}^{2}\widehat{v}_{jj})$ is
non-positive for all $T$. Hence, since all $\mu _{j}$ are positive by
supposition, as $T$ $\longrightarrow \infty $, by (\ref{q1}) we have that 
\begin{equation*}
Q_{1}\geq \max_{j\in A}\{\widehat{\Upsilon }_{j}\}\min_{j\in A}\{\sqrt{T}%
\widehat{\mu }_{j}\}.
\end{equation*}%
with probability tending to $1.$ Because the mapping from a positive
semi-definite matrix to its maximum eigenvalue is continuous on the space of
such matrices, by [D3] we have $\widehat{\lambda }_{\max }\overset{p}{%
\longrightarrow }\lambda _{\max }$ where $\lambda _{\max }$ is the largest
eigenvalue of $V$. By [D2], $0<\lambda _{\max }<\infty $ and thus we have 
\begin{equation*}
Q_{1}/Q_{2}\geq \min_{j\in A}\{\sqrt{T}\widehat{\mu }_{j}\}/\sqrt{p\widehat{%
\lambda }_{\max }}
\end{equation*}%
with probability tending to $1$ as $T$ $\longrightarrow \infty .$ Since $%
\sqrt{T}\widehat{\mu }_{j}$ goes to infinity as $T$ $\longrightarrow \infty $
for $j\in A$, it follows that $Q=\Phi (Q_{1}/Q_{2})\overset{p}{%
\longrightarrow }1.$

\subsection{Proof of Theorem 3\label{pf_nor_thm2}}

Since rejection of $H_{0}$ occurs if $Q<\alpha $ for the test statistic (\ref%
{q}), it suffices for consistency to show that under $H_{1}$, $Q_{2}$ goes
in probability to some positive constant and $Q_{1}$ goes to minus infinity
as $T\longrightarrow \infty $. By (\ref{psi_hat}) and Lemma \ref%
{Ch_nor_lemma}, the probability limit of $\widehat{\Psi }$ under $H_{1}$ is
the $p$ dimensional vector whose $j$th element is $[1\{\mu _{j}<0\}+\Psi
(0)1\{\mu _{j}=0\}]$. Therefore, by [D3] and [D4] 
\begin{equation*}
Q_{2}\equiv \sqrt{\widehat{\Psi }^{\prime }\widehat{\Delta }\widehat{V}%
\widehat{\Delta }\widehat{\Psi }}\overset{p}{\longrightarrow }\sqrt{d(\mu
)^{\prime }\Delta V\Delta d(\mu )},
\end{equation*}%
which is strictly positive by the regularity condition [D2]. On the other
hand, Lemma \ref{Ch_nor_lemma2} implies that $\sqrt{T}\Psi _{T}(\widehat{%
\theta }_{j}\widehat{\mu }_{j})\widehat{\theta }_{j}\widehat{\mu }_{j}$ is
bounded in probability for $j\in J\backslash B$ but tends to negative
infinity for $j\in B$. Furthermore, Lemma \ref{Lemma 3} implies that $%
\Lambda _{T}(\widehat{\theta }_{j}\widehat{\mu }_{j},\widehat{\theta }%
_{j}^{2}\widehat{v}_{jj})=o_{p}(1)$ for $j\in J$. Under $H_{1}$, $B$ is
non-empty and thus $Q_{1}/Q_{2}$ goes to $-\infty $ in probability and hence 
$P(Q<\alpha )\longrightarrow 1$ as $T\longrightarrow \infty $ .

\subsection{Proof of Theorem 4\label{pf_nor_thm3}}

Under the assumed form of local sequence (\ref{local}), for all $j$ we have%
\begin{equation*}
K(T)\widehat{\theta }_{j}\widehat{\mu }_{j}=(K(T)/\sqrt{T})\widehat{\theta }%
_{j}[\sqrt{T}(\widehat{\mu }_{j}-\mu _{j})+c_{j}]+K(T)\widehat{\theta }%
_{j}\gamma _{j}
\end{equation*}%
where $\gamma _{j}\geq 0$. In the case $\gamma _{j}=0$, Assumptions [A4],
[D1] and [D4] imply that $K(T)\widehat{\theta }_{j}\widehat{\mu }_{j}\overset%
{p}{\longrightarrow }$\textit{\ }$0$ as $T\longrightarrow \infty $ . By [A2]
and the continuous mapping theorem, this then implies that $\Psi (K(T)%
\widehat{\theta }_{j}\widehat{\mu }_{j})\overset{p}{\longrightarrow }\mathit{%
\ }\Psi (0).$ On the other hand, if $\gamma _{j}>0$, (\ref{local}) implies
that there is some $\delta >0$ such that $\mu _{j}>\gamma _{j}-\delta >0$
for all $T$ sufficiently large. So under [A1], [A3], [A6], [D1] and [D4], we
have that $\sqrt{T}\Psi _{T}(\widehat{\theta }_{j}\widehat{\mu }_{j})%
\widehat{\theta }_{j}\widehat{\mu }_{j}\overset{p}{\longrightarrow }$\textit{%
\ }$0$ by using arguments closely matching the proof of part (1) of Lemma %
\ref{Ch_nor_lemma}.\medskip

Therefore, from these results and by (\ref{local}), [D1], [D4] and Lemma \ref%
{Lemma 3}, $Q_{1}$ is asymptotically equivalent in probability to%
\begin{equation*}
\Psi (0)\dsum\limits_{j=1}^{p}1\{\gamma _{j}=0\}\theta _{j}[\sqrt{T}(%
\widehat{\mu }_{j}-\mu _{j})+c_{j}]
\end{equation*}%
and thus has an asymptotic normal distribution with mean $\Psi (0)\tau $ and
variance $\Psi (0)^{2}\kappa $. Using similar arguments, it is
straightforward to see that $Q_{2}\overset{p}{\longrightarrow }$\textit{\ }$%
\Psi (0)\sqrt{\kappa }$. Therefore, $Q_{1}/Q_{2}\overset{d}{\longrightarrow }%
N(\kappa ^{-1/2}\tau ,1)$ from which the assertion of Theorem \ref{local
power theorem} follows.

\subsection{Proof of Theorem 5\label{B7}}

We shall establish that for any non-zero vector $c$, 
\begin{equation}
\Phi (z_{\alpha }+\sqrt{c^{\prime }V^{-1}c})\geq P(S(Z+c,V)>q_{\alpha })
\label{local dominance}
\end{equation}%
holds for every testing function $S(.,.)$\ such that $P(S(Z,V)>q_{\alpha
})=\alpha $\ under $Z\sim N(0,V)$. The theorem then follows by noting that
the left-hand side of (\ref{local dominance}) when $c=-\delta V\theta $
coincides with the power function (\ref{nor_thm3}) under the local direction
specified by the theorem.\medskip

To show (\ref{local dominance}), consider an imaginary situation where $X$
is the observable random vector that is distributed as $Z+\mu _{X}$ where $%
Z\sim N(0,V)$. For given $V$, a simple application of the Neyman-Pearson
lemma (Lehmann and Romano (2005, p.\thinspace 60, Theorem 3.2.1)) implies
that a most powerful test at level $\alpha $ of the simple null hypothesis $%
\mu _{X}=0$ versus the simple alternative $\mu _{X}=c$ is to reject the null
if and only if $-c^{\prime }V^{-1}X/\sqrt{c^{\prime }V^{-1}c}<z_{\alpha }$.
Hence (\ref{local dominance}) holds by noting that such test has power equal
to $\Phi (z_{\alpha }+\sqrt{c^{\prime }V^{-1}c})$ which is therefore not
smaller than $P(S(Z+c,V)>q_{\alpha })$, the power of another test at level $%
\alpha $ which rejects the null hypothesis $\mu _{X}=0$ if and only if $%
S(X,V)>$ $q_{\alpha }$.

\subsection{Sufficient Condition for Assumption [U2]\label{B8}}

The following lemma provides a sufficient condition for Assumption [U2] of
Section \ref{nor_s4}. Recall that $Y\equiv \sqrt{T}(\widehat{\mu }-\mu ).$%
\bigskip

\begin{lemma}
\textit{Assumption [U2] holds provided that given any finite scalar }$c$%
\textit{, }%
\begin{equation}
\lim_{T\longrightarrow \infty }|P_{G_{T}}(\beta _{T}^{\prime }Y\leq c)-\Phi
(c)|\text{ }=0  \label{U2_s}
\end{equation}%
\textit{for any sequence }$(G_{T},\beta _{T})$\textit{\ satisfying }$%
G_{T}\in \Gamma _{0}$\textit{\ and }$\beta _{T}^{\prime }V_{G_{T}}\beta
_{T}=1$\textit{.}
\end{lemma}

\bigskip

\begin{proof}
Let%
\begin{equation*}
f_{T}(G,\beta )\equiv |P_{G}(\beta ^{\prime }Y\leq c)-\Phi (c)|\text{.}
\end{equation*}%
Let $S$ denote the set $\{(G,\beta ):G\in \Gamma _{0},\beta \in \Sigma (G)\}$
where the set $\Sigma (G)\equiv \{\beta \in R^{p}:\beta ^{\prime }V_{G}\beta
=1\}$. Note that 
\begin{equation}
\sup_{G\in \Gamma _{0}}\sup_{\beta \in \Sigma (G)}f_{T}(G,\beta
)=\sup_{(G,\beta )\in S}f_{T}(G,\beta ).  \label{fT}
\end{equation}%
Since for any $\varepsilon >0$, there is a pair $(G_{T}(\varepsilon ),\beta
_{T}(\varepsilon ))$ in $S$ such that 
\begin{equation*}
\sup_{(G,\beta )\in S}f_{T}(G,\beta )<f_{T}(G_{T}(\varepsilon ),\beta
_{T}(\varepsilon ))+\varepsilon ,
\end{equation*}%
Assumption (\ref{U2_s}) used with equality (\ref{fT}) implies 
\begin{equation*}
\lim_{T\longrightarrow \infty }\sup_{G\in \Gamma _{0}}\sup_{\beta \in \Sigma
(G)}f_{T}(G,\beta )<\varepsilon .
\end{equation*}%
Hence Assumption [U2] follows by noting that $\varepsilon $ is arbitrary
chosen and $f_{T}\geq 0$.
\end{proof}

\subsection{Proof of Theorem 2\label{B9}}

We aim to establish the inequality%
\begin{equation}
\limsup_{T\longrightarrow \infty }\sup_{G\in \Gamma _{0}}P_{G}(Q<\alpha
)\leq \alpha .  \label{uniform size 1}
\end{equation}%
Then Theorem \ref{uniform size convergence} follows by combining together
the results implied by (\ref{uniform size 1}) and Part (1) of Theorem \ref%
{ch_nor_thm1}.\medskip

Let $z_{\alpha }$ be the $\alpha $ quantile of the standard normal
distribution. The test rejects the null hypothesis if and only if $Q_{2}>0$
and $Q_{1}-z_{\alpha }Q_{2}<0$. Therefore, 
\begin{equation}
P_{G}(\text{reject }H_{0})\leq P_{G}(Q_{1}-z_{\alpha }Q_{2}<0).
\label{size bound}
\end{equation}%
The strategy of the proof is to demonstrate that $P_{G}(Q_{1}-z_{\alpha
}Q_{2}<0)$ is asymptotically bounded by the nominal size $\alpha $ uniformly
for all $G$ satisfying the null hypothesis. That then validates (\ref%
{uniform size 1}) via (\ref{size bound}). Note that $-z_{\alpha }>0$ for $%
0<\alpha <1/2$ as used in this theorem. By (\ref{q1}), (\ref{q2}) and
non-positivity of the $\Lambda _{T}$ term, we have

\begin{eqnarray*}
Q_{1} &\geq &\dsum\limits_{j=1}^{p}\Psi (K(T)\widehat{\theta }_{j}\widehat{%
\mu }_{j})\sqrt{T}\widehat{\theta }_{j}\widehat{\mu }_{j} \\
Q_{2} &=&\sqrt{\dsum\limits_{i=1}^{p}\dsum\limits_{j=1}^{p}\Psi (K(T)%
\widehat{\theta }_{i}\widehat{\mu }_{i})\Psi (K(T)\widehat{\theta }_{j}%
\widehat{\mu }_{j})\widehat{\theta }_{i}\widehat{\theta }_{j}\widehat{v}_{ij}%
}
\end{eqnarray*}%
where $\widehat{v}_{ij}$ and $v_{ij}$ are the $(i,j)$ elements of $\widehat{V%
}$ and $V_{G}$, respectively. For notational simplicity, the dependence of $%
\mu $ and $v_{ij}$ on $G$ is kept implicit. \medskip

Now we give details of the proof. For ease of presentation, they are
organized in the following headed subsections.

\subsubsection*{1. Lower Bound for the Difference $(Q_{1}-z_{\protect\alpha %
}Q_{2})$}

Let $\delta _{T}\equiv \sqrt{K(T)/\sqrt{T}}$. For any $\eta >0$, define the
set%
\begin{equation*}
R_{T}(\mu )\equiv \{j:0\leq K(T)\mu _{j}\leq 2\eta \delta _{T}\}.
\end{equation*}%
We show that, with probability tending to 1 uniformly over $G\in \Gamma _{0}$
as $T\longrightarrow \infty $,%
\begin{equation}
Q_{1}-z_{\alpha }Q_{2}\geq Q_{1,R_{T}}-z_{\alpha }Q_{2,R_{T}}
\label{uniform size 2}
\end{equation}%
where 
\begin{eqnarray*}
Q_{1,R_{T}} &\equiv &\dsum\limits_{j\in R_{T}(\mu )}^{{}}\Psi (K(T)\widehat{%
\theta }_{j}\widehat{\mu }_{j})\sqrt{T}\widehat{\theta }_{j}\widehat{\mu }%
_{j}, \\
Q_{2,R_{T}} &\equiv &\sqrt{\dsum\limits_{i\in R_{T}(\mu
)}^{{}}\dsum\limits_{j\in R_{T}(\mu )}^{{}}\Psi (K(T)\widehat{\theta }_{i}%
\widehat{\mu }_{i})\Psi (K(T)\widehat{\theta }_{j}\widehat{\mu }_{j})%
\widehat{\theta }_{i}\widehat{\theta }_{j}\widehat{v}_{ij}}.
\end{eqnarray*}%
We follow the convention that summation over an empty set yields value zero.
Note that (\ref{uniform size 2}) automatically holds when $R_{T}(\mu
)=\{1,2,...,p\}$. For $R_{T}(\mu )$ being a proper subset of $\{1,2,...,p\}$%
, we rely on the fact (proved in the next subsection) that, with probability
tending to 1 uniformly over $G\in \Gamma _{0}$ as $T\longrightarrow \infty $,%
\begin{equation}
K(T)\widehat{\mu }_{j}>\eta \delta _{T}\text{ for }j\notin R_{T}(\mu )
\label{uniform size 3}
\end{equation}%
and, for $R_{T}(\mu )$ nonempty, 
\begin{equation}
Q_{2,R_{T}}>\sqrt{\omega ^{\prime }/2}>0  \label{uniform size 4}
\end{equation}%
where $\omega ^{\prime }$ is the constant defined in Assumption [U4]-(ii).
Let $m$ be any index such that $m\notin R_{T}(\mu )$ and $\widehat{\theta }%
_{m}\widehat{\mu }_{m}\leq \widehat{\theta }_{j}\widehat{\mu }_{j}$ for all $%
j\notin R_{T}(\mu )$. Since $\Psi $ is non-negative, (\ref{uniform size 3})
implies%
\begin{equation}
Q_{1}\geq Q_{1,R_{T}}+\Psi (K(T)\widehat{\theta }_{m}\widehat{\mu }_{m})%
\widehat{\theta }_{m}\eta \delta _{T}^{-1}.  \label{uniform size 5}
\end{equation}%
Furthermore, by [A1] the function $\Psi $ is non-increasing and $\Psi \leq 1$%
. Thus, (\ref{uniform size 3}) and (\ref{uniform size 4}) together imply 
\begin{equation}
\left\vert Q_{2,R_{T}}-Q_{2}\right\vert \leq \left\vert
Q_{2,R_{T}}^{2}-Q_{2}^{2}\right\vert /Q_{2,R_{T}}\leq p^{2}\Psi (K(T)%
\widehat{\theta }_{m}\widehat{\mu }_{m})\left\Vert \widehat{\Delta }%
\right\Vert ^{2}\left\Vert \widehat{V}\right\Vert \sqrt{2/\omega ^{\prime }}.
\label{uniform size 6}
\end{equation}

Given that $-z_{\alpha }>0$, when $R_{T}(\mu )$ is empty, (\ref{uniform size
5}) alone implies (\ref{uniform size 2}). With $R_{T}(\mu )$ non-empty, (\ref%
{uniform size 5}) and (\ref{uniform size 6}) together imply (\ref{uniform
size 2}) provided%
\begin{equation}
\widehat{\theta }_{m}\eta \delta _{T}^{-1}\geq -z_{\alpha }p^{2}\left\Vert 
\widehat{\Delta }\right\Vert ^{2}\left\Vert \widehat{V}\right\Vert \sqrt{%
2/\omega ^{\prime }}.  \label{uniform size 7}
\end{equation}%
We show that under the null hypothesis, (\ref{uniform size 3}), (\ref%
{uniform size 4}) and (\ref{uniform size 7}) will indeed hold for $\eta $
small enough and $T$ large enough (yielding $\delta _{T}$ small enough by
Assumption [A4]) under the key event $E_{T}^{\eta }$ described next.

\subsubsection*{2. The Key Event $E_{T}^{\protect\eta }$ and Lower Bound for
the Difference $(Q_{1,R_{T}}-z_{\protect\alpha }Q_{2,R_{T}})$}

Let $Y_{j}$ be the $j$th element of $Y\equiv \sqrt{T}(\widehat{\mu }-\mu )$.
For $\eta >0$, define the event%
\begin{equation*}
E_{T}^{\eta }\equiv \{\delta _{T}\left\Vert Y\right\Vert <\eta ,||\widehat{V}%
-V_{G}||\text{ }<\eta ,\left\Vert \widehat{\Delta }-\Delta \right\Vert <\eta
\delta _{T}\}
\end{equation*}%
which holds with probability tending to 1 uniformly over $G\in \Gamma _{0}$
as $T\longrightarrow \infty $ by Assumptions [A4], [U1] and [U3]-(ii). Since 
$K(T)\widehat{\mu }_{j}=K(T)\mu _{j}+\delta _{T}^{2}Y_{j}$, under the null
hypothesis the event $E_{T}^{\eta }$ implies the inequality (\ref{uniform
size 3}). To show that the event $E_{T}^{\eta }$ also implies (\ref{uniform
size 4}) and (\ref{uniform size 7}), and then derive the key result (\ref%
{uniform size 8}) of this subsection, we first need to draw out the
following inequalities (\ref{e1}) - (\ref{q2r_}).\medskip

Note that when $0\leq K(T)\mu _{j}\leq 2\eta \delta _{T}$, we have that by
Assumption [U3]-(i) and under the event $E_{T}^{\eta }$,%
\begin{eqnarray}
\sqrt{T}\widehat{\theta }_{j}\widehat{\mu }_{j} &\geq &\theta
_{j}Y_{j}-3\eta ^{2},  \label{e1} \\
\left\vert K(T)\widehat{\theta }_{j}\widehat{\mu }_{j}\right\vert &\leq
&3\eta \delta _{T}(\lambda +\eta \delta _{T}).  \label{e2}
\end{eqnarray}%
By Assumption [A2], $\Psi (x)$ is differentiable on $\left\vert x\right\vert
\leq 3\eta \delta _{T}(\lambda +\eta \delta _{T})$ for $\eta $ small enough
and $T$ \ large enough.\medskip\ Therefore, given $\Psi \leq 1$, the event $%
E_{T}^{\eta }$ and inequalities (\ref{e1}) and (\ref{e2}) imply that 
\begin{equation*}
\Psi (K(T)\widehat{\theta }_{j}\widehat{\mu }_{j})\sqrt{T}\widehat{\theta }%
_{j}\widehat{\mu }_{j}\geq \Psi (0)\theta _{j}Y_{j}-3(\lambda b_{\Psi
}(\lambda +\eta \delta _{T})+1)\eta ^{2}
\end{equation*}%
where $b_{\Psi }$ denotes the bound on the derivative of $\Psi (x)$ defined
in Assumption [A2]. Hence, when $\eta <1$ and $\delta _{T}<1$, we may
certainly write%
\begin{equation}
Q_{1,R_{T}}\geq \Psi (0)\dsum\limits_{j\in R_{T}(\mu )}^{{}}\theta
_{j}Y_{j}-C_{1}\eta  \label{q1r}
\end{equation}%
where $C_{1}$ is a fixed positive quantity given values of $p$, $\lambda $
and $b_{\Psi }$. By Assumptions [U3]-(i) and [U4]-(i) and using similar
arguments with $\eta <1$ and $\delta _{T}<1$, we can obtain a bound for $%
Q_{2,R_{T}}^{2}$ under the event $E_{T}^{\eta }$ as the following%
\begin{equation}
Q_{2,R_{T}}^{2}\geq \Psi (0)^{2}\dsum\limits_{i\in R_{T}(\mu
)}^{{}}\dsum\limits_{j\in R_{T}(\mu )}^{{}}\theta _{i}\theta
_{j}v_{ij}-C_{2}\eta  \label{q2r_}
\end{equation}%
where $C_{2}$ is fixed and positive given values of $p$, $\lambda $, $\omega 
$, $b_{\Psi }$ and $\Psi (0)$. \medskip

We can choose $\eta $ to satisfy $\eta <\min \{1,\omega ^{\prime
}/(2C_{2})\} $ and choose $T$ such that $2\eta \delta _{T}/K(T)<\sigma $,
where $\sigma $ is the constant defined in Assumption [U4] by which the
right-hand side of (\ref{q2r_}) is larger than $\omega ^{\prime }/2$ and
hence inequality (\ref{uniform size 4}) is satisfied. Using Assumptions
[U3]-(i) and [U4]-(i), under the event $E_{T}^{\eta }$, we see $\widehat{%
\theta }_{m}>\lambda ^{\prime }-\delta _{T}\eta $ whilst $\left\Vert 
\widehat{\Delta }\right\Vert ^{2}\left\Vert \widehat{V}\right\Vert \leq
(\lambda +\delta _{T}\eta )^{2}(\omega +\eta )$. Since $\delta
_{T}^{-1}\longrightarrow \infty $ by Assumption [A4], given $\eta >0$, (\ref%
{uniform size 7}) will indeed hold for large enough $T$ . Finally, let $%
r_{T} $ denote the $p$ dimensional vector whose $j$th element is $\theta
_{j} $ if $j\in R_{T}(\mu )$ and zero, otherwise. Then given that $%
-z_{\alpha }>0$ and with $\eta $ small enough and $T$ large enough, (\ref%
{q1r}) and (\ref{q2r_}) together imply%
\begin{equation}
Q_{1,R_{T}}-z_{\alpha }Q_{2,R_{T}}\geq \Psi (0)r_{T}^{\prime }Y-C_{1}\eta
-z_{\alpha }\sqrt{\Psi (0)^{2}r_{T}^{\prime }V_{G}r_{T}-C_{2}\eta }.
\label{uniform size 8}
\end{equation}

\subsubsection*{3. The Probability Bounds}

We have shown above how occurrence of the event $E_{T}^{\eta }$ implies the
inequality (\ref{uniform size 2}) given $\eta $ small enough and $T$ large
enough. Hence 
\begin{eqnarray}
P_{G}(Q_{1}-z_{\alpha }Q_{2}<0) &\leq &1-P_{G}(E_{T}^{\eta
})+P_{G}(Q_{1}-z_{\alpha }Q_{2}<0,E_{T}^{\eta })  \notag \\
&\leq &1-P_{G}(E_{T}^{\eta })+P_{G}(Q_{1,R_{T}}-z_{\alpha }Q_{2,R_{T}}<0)
\label{uniform size 9}
\end{eqnarray}%
where the last term of (\ref{uniform size 9}) is zero when $R_{T}(\mu )$ is
empty. For non-empty $R_{T}(\mu )$, using (\ref{uniform size 8}) yields 
\begin{equation}
P_{G}(Q_{1,R_{T}}-z_{\alpha }Q_{2,R_{T}}<0)\leq P_{G}(r_{T}^{\prime
}Y-z_{\alpha }\sqrt{r_{T}^{\prime }V_{G}r_{T}-C_{2}\eta /\Psi (0)^{2}}%
<C_{1}\eta /\Psi (0)).  \label{uniform size 10}
\end{equation}%
The probability in the right-hand side of (\ref{uniform size 10}) may be
written as

\begin{equation}
P_{G}(\beta _{T}^{\prime }Y<z_{\alpha }\widetilde{C}_{2,R_{T}}+\eta 
\widetilde{C}_{1,R_{T}})  \label{uniform size 11}
\end{equation}%
where%
\begin{eqnarray*}
\beta _{T} &\equiv &r_{T}/\sqrt{r_{T}^{\prime }V_{G}r_{T}}, \\
\widetilde{C}_{1,R_{T}} &\equiv &C_{1}/(\Psi (0)\sqrt{r_{T}^{\prime
}V_{G}r_{T}}), \\
\widetilde{C}_{2,R_{T}} &\equiv &\sqrt{r_{T}^{\prime }V_{G}r_{T}-C_{2}\eta
/\Psi (0)^{2}}/\sqrt{r_{T}^{\prime }V_{G}r_{T}}.
\end{eqnarray*}%
Note that by [U4]-(ii), we have that with $T$ large enough, $0\leq 
\widetilde{C}_{1,R_{T}}\leq C_{1}/(\Psi (0)\sqrt{\omega ^{\prime }})$ and $%
\sqrt{1-C_{2}\eta /\omega ^{\prime }}\leq \widetilde{C}_{2,R_{T}}\leq 1$.
Hence, given $z_{\alpha }<0$ and $\eta $ small enough, the probability (\ref%
{uniform size 11}) cannot exceed 
\begin{equation}
P_{G}(\beta _{T}^{\prime }Y<z_{\alpha }\sqrt{1-C_{2}\eta /\omega ^{\prime }}%
+C_{1}\eta /(\Psi (0)\sqrt{\omega ^{\prime }})).  \label{uniform size 12}
\end{equation}

Given the fact that $\beta _{T}$ is non-stochastic with $\beta _{T}^{\prime
}V_{G}\beta _{T}=1$, Assumption [U2] implies that given $\eta $, for any $%
\xi >0,$ there is a threshold $T^{\ast }(\eta ,\xi )$ such that for $%
T>T^{\ast }(\eta ,\xi )$, the probability (\ref{uniform size 12}) will be
smaller than%
\begin{equation*}
\Phi (z_{\alpha }\sqrt{1-C_{2}\eta /\omega ^{\prime }}+C_{1}\eta /(\Psi (0)%
\sqrt{\omega ^{\prime }}))+\xi
\end{equation*}%
uniformly over all $G$ obeying the null hypothesis. On the other hand, by
Assumptions [A4], [U1] and [U3]-(ii) applied to the event $E_{T}^{\eta }$,
for any $\varepsilon >0$, there is a threshold $T^{\ast \ast }(\eta
,\varepsilon )$ such that for $T>T^{\ast \ast }(\eta ,\varepsilon )$, $%
P_{G}(E_{T}^{\eta })>1-\varepsilon $ uniformly over all $G$ obeying the null
hypothesis. Putting together these facts and (\ref{uniform size 9}), (\ref%
{uniform size 10}), (\ref{uniform size 12}), we have that for $T>\max
\{T^{\ast }(\eta ,\xi ),T^{\ast \ast }(\eta ,\varepsilon )\}$, 
\begin{equation*}
\sup_{G\in \Gamma _{0}}P_{G}(Q_{1}-z_{\alpha }Q_{2}<0)\leq \Phi (z_{\alpha }%
\sqrt{1-C_{2}\eta /\omega ^{\prime }}+C_{1}\eta /(\Psi (0)\sqrt{\omega
^{\prime }}))+\xi +\varepsilon
\end{equation*}%
from which by letting $T\longrightarrow \infty $ in accordance with $T>\max
\{T^{\ast }(\eta ,\xi ),T^{\ast \ast }(\eta ,\varepsilon )\}$ as the scalars 
$\eta $, $\xi $ and $\varepsilon $ approach zero, it follows that $%
\limsup_{T\longrightarrow \infty }\sup_{G\in \Gamma
_{0}}P_{G}(Q_{1}-z_{\alpha }Q_{2}<0)\leq \alpha $.

\section{Covariance Singularity Examples\label{nor_regularity condition}}

In this appendix section, we present three examples of estimator covariance
singularity for which the high level assumptions [D2] and [U4]-(ii) are
verified. Recall that $G$ is the joint distribution from which the
underlying individual data vector is randomly sampled. $\Gamma $ is the set
of all possible $G$ compatible with presumed specification of the data
generating process and $\Gamma _{0}$ is the subset of $\Gamma $ that
satisfies the null hypothesis. All parameter values such as $\mu $ and $V$
depend on the point $G$ of evaluation but we keep that implicit to avoid
notational clutter.\medskip

In the first two examples, the econometric model is initially characterized
by an $r$ dimensional vector of parameters $\beta \equiv (\beta _{1},\beta
_{2},...,\beta _{r})^{\prime }$. The restrictions being tested are
synthesized into the one-sided form $\mu \geq 0$ with $\mu =(\mu _{1},\mu
_{2},...,\mu _{p})^{\prime }=C\beta +b$ where $C$ is a known $p\times r$
matrix and $b$ is a known $p$ dimensional vector of constants.\ We assume an
asymptotically normal estimator $\widehat{\beta }$ is available with
non-singular asymptotic variance matrix $\Omega $. Since $V=C\Omega
C^{\prime }$, $V$ value induced by any $G\in \Gamma $ is necessarily
singular when $r<p$. In the third example, we consider a different scenario
where singularity arises only for some specific $V$ values.

\subsection*{Example 1: Triangle Restriction\protect\smallskip \label%
{production example}}

For a Cobb-Douglas production function with capital and labor elasticity
coefficients $\beta _{1}$ and $\beta _{2}$, the restrictions being tested $%
\beta _{1}\geq 0,$ $\beta _{2}\geq 0$ and $\beta _{1}+\beta _{2}\leq 1$
(non-increasing returns to scale) form a triangle for the graph of $(\beta
_{1},\beta _{2})$. Here $r=2$, $p=3$ and 
\begin{equation}
\mu =(\mu _{1},\mu _{2},\mu _{3})^{\prime }=(\beta _{1},\beta _{2},1-\beta
_{1}-\beta _{2})^{\prime }.  \label{example 1}
\end{equation}

\textbf{Verification of [D2] and [U4]-(ii) : }Note that $V=C\Omega C^{\prime
}$ where $\Omega $ is the variance matrix of the asymptotic distribution of $%
\sqrt{T}($ $\widehat{\beta }-\beta )$ and%
\begin{equation*}
C^{\prime }=\left[ 
\begin{array}{ccc}
1 & 0 & -1 \\ 
0 & 1 & -1%
\end{array}%
\right] \text{, \ \ }C^{\prime }\Delta d(\mu )=\left[ 
\begin{array}{ccc}
\theta _{1} & 0 & -\theta _{3} \\ 
0 & \theta _{2} & -\theta _{3}%
\end{array}%
\right] d(\mu ).
\end{equation*}%
We assume the primitive condition that the smallest eigenvalue of $\Omega $
is bounded away from zero over all $G\in \Gamma $. Assumption [D2] is true
since $C^{\prime }\Delta d(\mu )$ being zero for non-zero $d(\mu )$ would
require all elements of $d(\mu )$ to be non-zero, in turn requiring all
elements of $\mu $ given by (\ref{example 1}) to be negative or zero, which
is impossible. For Assumption [U4]-(ii), we note that for sufficiently small 
$\sigma $, the only non-zero values for $d_{\sigma }(\mu )$ possible under
the null hypothesis are $\Psi (0)$ multiples of $(1,0,0)^{\prime }$, $%
(0,1,0)^{\prime }$, $(0,0,1)^{\prime }$, $(1,1,0)^{\prime }$, $%
(1,0,1)^{\prime }$, $(0,1,1)^{\prime }$, because it is not possible for more
than two of the elements of $\mu $ to simultaneously lie between 0 and $%
\sigma <1/3$ as $\mu _{1}+\mu _{2}+\mu _{3}=1$. Therefore, given Assumption
[U3]-(i) and the primitive condition on $\Omega $, Assumption [U4]-(ii) is
satisfied here.

\subsection*{Example 2: Interval Restrictions\protect\smallskip\ with Fixed
Known\ End-Points\label{interval restriction}}

Suppose the$\ r$ dimensional parameter vector $\beta $ is hypothesized to
satisfy interval restrictions $l\leq \beta \leq u,\ $where $l$ and $u$ are
numerically specified. In this case, $p=2r$ and $\mu =((\beta -l)^{\prime
},(u-\beta )^{\prime })^{\prime }$. An estimator $\widehat{\beta }$ is
available such that $\sqrt{T}($ $\widehat{\beta }-\beta )$ is asymptotically
normal with variance $\Omega $ whose smallest eigenvalue is assumed
primitively to be bounded away from zero over all $G\in \Gamma $. Note that $%
V=C\Omega C^{\prime }$ where $C^{\prime }=[I_{r},-I_{r}]$. Thus, $C^{\prime
}\Delta d(\mu )$ is the $r$ dimensional vector whose $j$th element is 
\begin{equation}
\lbrack 1\{\beta _{j}<l_{j}\}+\Psi (0)1\{\beta _{j}=l_{j}\}]\theta
_{j}-[1\{\beta _{j}>u_{j}\}+\Psi (0)1\{\beta _{j}=u_{j}\}]\theta _{j+r}
\label{verify interval}
\end{equation}%
for $j\leq r$. We consider the following two cases of interval hypotheses.

\subsubsection*{Case I : All hypothesized intervals are non-degenerate}

For Case I, the null hypothesis concerns only non-degenerate intervals in
the sense that $u_{j}>l_{j}$ for all $j\leq r$. \medskip

\textbf{Verification of [D2] and [U4]-(ii) for null hypothesis given by Case
I : }Note that under $H_{1}$, $\beta _{j}<l_{j}$ or $\beta _{j}>u_{j}$ for
some $j\leq r$ and thus (\ref{verify interval}) is either $\theta _{j}$ or $%
-\theta _{j+r}$ for some $j\leq r$. Hence $C^{\prime }\Delta d(\mu )$ is
non-zero and Assumption [D2] holds under the alternative hypothesis. We need
to further show that $C^{\prime }\Delta d(\mu )$ is not equal to zero for
non-zero $d(\mu )$ under the null hypothesis. But under $H_{0}$, (\ref%
{verify interval}) simplifies to%
\begin{equation}
\Psi (0)\left[ 1\{\beta _{j}=l_{j}\}\theta _{j}-1\{\beta _{j}=u_{j}\}\theta
_{j+r}\right] .  \label{verify 2}
\end{equation}%
for all $j\leq r$. Given that $u_{j}>l_{j}$ for all $j$, there is some $j$
such that expression (\ref{verify 2}) equals either $\Psi (0)\theta _{j}$ or 
$-\Psi (0)\theta _{j+r}$ whenever $d(\mu )$ is non-zero under the null
hypothesis. Hence, Assumption [D2] is verified. \medskip

We now verify the high level assumption [U4]-(ii). Under the null
hypothesis, the $j$th element of $C^{\prime }\Delta d_{\sigma }(\mu )$ is 
\begin{equation}
\Psi (0)[1\{l_{j}+\sigma \geq \beta _{j}\geq l_{j}\}\theta _{j}-1\{u_{j}\geq
\beta _{j}\geq u_{j}-\sigma \}\theta _{j+r}].  \label{verify 3}
\end{equation}%
For $\sigma <\min_{j\in \{1,2,...,r\}}(u_{j}-l_{j})/2$, if $d_{\sigma }(\mu
) $ is a non-zero, then there is some $j$ such that expression (\ref{verify
3}) equals either $\Psi (0)\theta _{j}$ or $-\Psi (0)\theta _{j+r}$ and thus 
$C^{\prime }\Delta d_{\sigma }(\mu )$ is a non-zero vector of length which
is bounded away from zero by Assumption [U3]-(i). Given the primitive
eigenvalue assumption on $\Omega $, this completes verification of
Assumption [U4]-(ii).

\subsubsection*{Case II : At least one hypothesized interval is degenerate}

For Case II, at least one interval is specified to be degenerate (i.e. $%
l_{j}=u_{j}$ for some $j\leq r$) in the null hypothesis. Let $S_{e}$ denote
the subset of $\{1,2,...,r\}$ such that $l_{j}=u_{j}$ holds for all $j\in
S_{e}$ but $l_{j}<u_{j}$ for all $j\notin S_{e}$. \medskip

\textbf{Verification of [D2] and [U4]-(ii) for null hypothesis given by Case
II : } Under $H_{1}$, Assumption [D2] holds by the same arguments as given
in Case I. Under $H_{0}$, (\ref{verify 2}) becomes $\Psi (0)\left( \theta
_{j}-\theta _{j+r}\right) $ for all $j\in S_{e}$. In this case, Assumption
[D2] still holds but the restriction that $\theta _{j}\neq \theta _{j+r}$
for at least one $j\in S_{e}$ has to be imposed. This extra restriction
guarantees that $C^{\prime }\Delta d(\mu )$ is not equal to zero for all
non-zero $d(\mu )$ and thus [D2] is fulfilled. \medskip

We now verify the high level assumption [U4]-(ii). Note that [U4]-(ii) only
concerns the null hypothesis under which (\ref{verify 3}) becomes $\Psi
(0)\left( \theta _{j}-\theta _{j+r}\right) $ for all $j\in S_{e}$.
Therefore, provided that there is one $j\in S_{e}$ such that $\left\vert
\theta _{j}-\theta _{j+r}\right\vert $ is bounded away from zero over all $%
G\in \Gamma _{0}$, then $C^{\prime }\Delta d_{\sigma }(\mu )$ is also a
non-zero vector of length which is bounded away from zero. Given the
primitive condition on $\Omega $, Assumption [U4]-(ii) is thus satisfied for
any $\sigma >0$.\medskip

We now comment on testing interval hypothesis of the Case II type within the
framework of this paper. For validity of the test, it suffices to choose any
single equality hypothesis indexed by $h\in S_{e}$ and specify $\theta
_{h}\neq \theta _{h+r}$ at the outset. This single asymmetry requirement is
the only operational difference compared with Case I. Moreover, since $%
v_{h,h}=v_{h+r,h+r}$ where $v_{h,h}$ denotes the $h$-th diagonal element of $%
V$, weighting inversely proportional to standard error is not ruled out. The
user can indeed set%
\begin{equation}
\theta _{h+r}=(1+\varepsilon )\theta _{h}\text{ with }\theta _{h}=1/\sqrt{%
v_{h,h}}\text{, }\varepsilon >-1\text{ and }\varepsilon \neq 0.
\label{adjustment}
\end{equation}%
Here $\varepsilon $ is a non-stochastic quantity chosen by the user to
control the degree of deviation from perfect standardization of the estimate 
$\widehat{\mu }_{h+r}$. The weighting scheme (\ref{adjustment}) ensures that
the test has exact asymptotic size in the uniform sense and is consistent
against all fixed alternatives. On the other hand, Theorem \ref{local power
theorem} suggests that the user can specify $\varepsilon <0$ (or reverse) to
attach more (or less) weight to detection of violation of $H_{0}$ in the
direction of $\beta _{h}<l_{h}$.\medskip

Note that asymmetric weighting (\ref{adjustment}) adopted here can be viewed
as \textquotedblleft perturbing\textquotedblright\ both $Q_{1}$ and $Q_{2}$
from the values they would take under symmetry. One might think to perturb
only $Q_{2}$ to ensure that singularity does not cause division by (near)
zero. For example, one could perturb $\widehat{V}$ in the expression (\ref%
{q2}) defining $Q_{2}$ in a manner akin to Andrews and Barwick (2012) who
adjust the $QLR$ test statistic by perturbing $\widehat{V}$ with a diagonal
matrix when the determinant of the correlation matrix induced by $\widehat{V}
$ is smaller than some pre-specified threshold. This alternative approach
can allow for symmetric weighting. However unperturbed $Q_{1}$ will
asymptotically converge to zero and hence rejection probability will tend to
zero under the null and local alternative scenarios where all non-degenerate
interval inequalities are non-binding. By contrast, the procedure (\ref%
{adjustment}) perturbing both $Q_{1}$ and $Q_{2}$ in a balanced way ensures
that the ratio $Q_{1}/Q_{2}$ stays asymptotically standard normal in the
null even when the only binding constraints are the equality hypotheses. It
thus enables non-zero test power to be retained in the aforementioned
scenarios of local alternatives.

\subsection*{Example 3: Interval Restrictions\protect\smallskip\ with
Unknown End-Points}

In Example 2, testing the inequalities $l\leq \beta \leq u$ was performed on
fixed known interval end-points. Suppose now that $l$ and $u$ are not known
but are parameters which satisfy $l\leq u$ and can take a continuum of
values including those which make $(u-l)$ arbitrarily close to zero as well
as precisely zero. There is no point estimator for $\beta $ but consistent
estimators $\widehat{l}$ and $\widehat{u}$ are available having joint
asymptotic normal distribution with variance matrix $\Omega $. This, for the
univariate case, is the scenario considered by Imbens and Manski (2004) and
Stoye (2009). For clarity, we stay with the setup where $\beta $ is a
scalar. We consider testing $H_{0}:l\leq \beta _{0}\leq u$ for a numerically
specified candidate value $\beta _{0}$ for $\beta $. We then take $\widehat{%
\mu }=(\beta _{0}-\widehat{l},\widehat{u}-\beta _{0})^{\prime }$ and $\mu
=(\beta _{0}-l,u-\beta _{0})^{\prime }$. The asymptotic distribution of $%
\sqrt{T}(\widehat{\mu }-\mu )$ is normal with variance%
\begin{equation*}
V=\left[ 
\begin{array}{cc}
\Omega _{11} & -\Omega _{12} \\ 
-\Omega _{12} & \Omega _{22}%
\end{array}%
\right] .
\end{equation*}

For any given $l$ and $u$, there is no reason why $V$ should be singular.
However, Stoye (2009. p. 1304, Lemma 3) demonstrates that, if one insists on 
$P(\widehat{u}\geq \widehat{l})=1$ holding over the underlying data
generating distribution space where the difference $(u-l)$ is bounded away
from infinity and the elements $\Omega _{11}$ and $\Omega _{22}$ bounded
away from zero and infinity, then $V$ necessarily depends on $(u-l)$ in such
a way that $\Omega _{12}-\Omega _{11}\longrightarrow 0$ and $\Omega
_{22}-\Omega _{11}\longrightarrow 0$ as $u-l\longrightarrow 0$. Thus,
singularity of $V$ where $\Omega _{11}=\Omega _{22}=\Omega _{12}$ must be
allowed for.\medskip

\textbf{Verification of [D2] and [U4]-(ii)} : For Assumption [D2], note that
under the maintained assumption that $l\leq u$, the vector $d(\mu )$ can be
non-zero only if it takes one of the following forms: $(1,0)^{\prime }$, $%
(0,1)^{\prime }$, $(\Psi (0),0)^{\prime }$, $(0,\Psi (0))^{\prime }$, $(\Psi
(0),\Psi (0))^{\prime }$. The first four of these cannot make $V\Delta d(\mu
)=0$. The last form can only occur when $l=\beta _{0}=u$ in which case we
have%
\begin{equation}
V\Delta d(\mu )=\Psi (0)[\theta _{1}\Omega _{11}-\theta _{2}\Omega
_{12},-\theta _{1}\Omega _{12}+\theta _{2}\Omega _{22}]^{\prime }.
\label{verification}
\end{equation}%
Note that (\ref{verification}) is zero only if $V$ is singular and $\theta
_{1}/\theta _{2}=\Omega _{12}/\Omega _{11}=\Omega _{22}/\Omega _{12}$.
Singularity occurs in Stoye's scenario where the model allows for $\Omega
_{11}=\Omega _{22}=\Omega _{12}$. Since the weights $\theta _{1}$ and $%
\theta _{2}$ are chosen by the user, we can use $\theta _{1}=1/\sqrt{\Omega
_{11}}$ and $\theta _{2}=(1+\varepsilon )/\sqrt{\Omega _{22}}$ where $%
\varepsilon $ is a pre-specified non-stochastic and non-zero quantity
satisfying $\varepsilon >-1$. Then Assumptions [D2] holds regardless of
singularity of $V$. For Assumption [U4]-(ii), we only need to consider the
null hypothesis. In this case, the possible forms of non-zero $d_{\sigma
}(\mu )$ can take are $(\Psi (0),0)^{\prime }$, $(0,\Psi (0))^{\prime }$ and 
$(\Psi (0),\Psi (0))^{\prime }$. It is easily seen that $d_{\sigma }(\mu
)^{\prime }\Delta V\Delta d_{\sigma }(\mu )$ equals $\Psi (0)^{2}$ for the
first, $\Psi (0)^{2}(1+\varepsilon )^{2}$ for the second, and $\Psi (0)^{2}%
\left[ \varepsilon ^{2}+2(1+\varepsilon )(1-\Omega _{12}/\sqrt{\Omega
_{11}\Omega _{22}})\right] $ for the third form. Hence Assumption [U4]-(ii)
holds. \medskip

In this example, the weights $\theta _{1}$ and $\theta _{2}$ are chosenly
asymmetrically and setting $\varepsilon $ to be greater (smaller) than zero
amounts to attaching more (or less) weight to detection of violation of $%
H_{0}$ in the direction $u<\beta _{0}$. The $\varepsilon $-perturbation
arguments adopted here are indeed based on those given in Case II of Example
2. The value of the perturbation parameter $\varepsilon $ is a user's input
to the test procedure. The choice does not affect validity of the results
concerning asymptotic test size and consistency. Asymmetry does affect local
power but, by the same device, offers the user an opportunity to input a
subjective assessment of the relative importance of different directions of
violation of the null hypothesis.

\end{document}